\documentclass[conference]{IEEEtran}
\IEEEoverridecommandlockouts

\usepackage{cite}
\usepackage{amsmath,amssymb,amsfonts}
\usepackage{algorithmic}
\usepackage{graphicx}
\usepackage{booktabs}
\usepackage{multirow}
\usepackage{textcomp}
\usepackage{xcolor}
\usepackage{tikz}
\usetikzlibrary{arrows.meta,positioning,decorations.pathreplacing,calc,fit}
\usepackage{pgfplots}
\pgfplotsset{compat=1.18}
\usepackage{url}
\usepackage[hidelinks,breaklinks]{hyperref}

\newcommand{\ket}[1]{\lvert #1\rangle}

\newcommand{\qy}{\textsc{Qyantram}}

\newcommand{\fitcol}[1]{%
  \resizebox{\ifdim\width>\columnwidth \columnwidth\else\width\fi}{!}{#1}}
\newcommand{\fitpage}[1]{%
  \resizebox{\ifdim\width>\textwidth \textwidth\else\width\fi}{!}{#1}}

\newcommand{\authorOne}{Sadique Khatib}
\newcommand{\authorTwo}{Sharyu Kekane}
\newcommand{\authorThree}{Abhushan Bokade}
\newcommand{\authorFour}{Dr.~Uday Wad}
\newcommand{\dept}{Department of Computer Science and Design}
\newcommand{\inst}{K.~K.~Wagh Institute of Engineering Education and Research}
\newcommand{\where}{Nashik 422003, India}

\newcommand{\deptshort}{Dept. of Computer Science and Design}
\newcommand{\instA}{K.~K.~Wagh Institute of Engineering}
\newcommand{\instB}{Education and Research}

\newcommand{\authorFive}{Dr.~S.~M.~Kamalapur}
\newcommand{\authorSix}{Dr.~Y.~D.~Bhise}

\newcommand{\contact}{spkhatib371323@kkwagh.edu.in}  
\newcommand{\emailTwo}{sharyukekane@gmail.com}       
\newcommand{\emailThree}{abhushan4625@gmail.com}     
\newcommand{\emailFour}{uday.wad@gmail.com}          
\newcommand{\emailFive}{smkamalapur@kkwagh.edu.in}   
\newcommand{\emailSix}{ydbhise@kkwagh.edu.in}        

\newcommand{\deptCE}{Dept. of Computer Engineering}

\begin{document}

\title{Qyantram: A Transparent and Verifiable\\Quantum Computing Platform}


%
%
\newcommand{\authblock}[2]{%
  \begin{tabular}[t]{@{}c@{}}
    \normalsize #1\\[0.55ex]
    \footnotesize \deptshort\\
    \footnotesize \instA\\
    \footnotesize \instB\\
    \footnotesize \where\\
    \footnotesize #2
  \end{tabular}}

\newcommand{\facultyblock}[4]{%
  \begin{tabular}[t]{@{}c@{}}
    \normalsize #1\\[0.55ex]
    \footnotesize \textit{#2}\\
    \footnotesize #3\\
    \footnotesize \instA\\
    \footnotesize \instB\\
    \footnotesize \where\\
    \footnotesize #4
  \end{tabular}}

\author{%
\begin{tabular*}{\textwidth}{@{\extracolsep{\fill}}ccc@{}}
  \authblock{\authorOne}{\contact}      &
  \authblock{\authorTwo}{\emailTwo}     &
  \authblock{\authorThree}{\emailThree} \\[3ex]
  \rule{0pt}{4.5ex}%
  \facultyblock{\authorFour}{Professor of Practice}{\deptCE}{\emailFour} &
  \facultyblock{\authorFive}{Head of Department}{\deptCE}{\emailFive}    &
  \facultyblock{\authorSix}{Departmental Coordinator}{\deptshort}{\emailSix}
\end{tabular*}}

%

\maketitle

\begin{abstract}
Most platforms dealing with quantum circuits only go as far as idealized
circuit assembly or simulation. The real engineering --- like compiling for
specific hardware, locating errors without measuring the qubit, or training
variational models --- usually gets hidden behind black-box calls. \qy{} is
built differently: it is an open platform, broken into four subsystems, and
everything revolves around a single rule --- every result comes bundled with
its supporting evidence, and no result gets through without an independent
double-check. The compilation pipeline is split into clear stages where each
one changes just one thing at a time, so if the pipeline adds a gate it is
easy to trace back and see exactly why it got added. The router publishes the
list of SWAP candidates that did not make it along with their scores, so
routing is not a guessing game. For verification, \qy{} focuses on the active
part of the layout: any qubit line left untouched stays locked at $\ket{0}$,
so the check only has to touch the occupied wires. The cost then scales as
$2^{k}$ for $k$ used qubits rather than $2^{N}$ for the whole device, and the
reduction is mathematically exact, not an approximation. That makes it
practical to run full per-compilation verification on 127- and 156-qubit
layouts, with speed-ups of around four orders of magnitude and the restricted
check holding steady below a millisecond. Error identification does not cut corners: every
possible pattern of Pauli errors gets checked, with no sampling, agreeing with
closed-form predictions to $2.8\times10^{-17}$. The machine-learning engine
lands within $2.0\times10^{-15}$ of \texttt{qiskit.quantum\_info} and trains
roughly $25$ to $50$ times faster. For full disclosure, the router adds about
$1.27\times$ as many SWAPs as Qiskit's SABRE because it skips optimisation ---
that extra cost is the price of keeping every routing decision visible from
start to finish.
\end{abstract}


\section{Introduction}
\label{sec:intro}

A quantum circuit as written and the same circuit as executed are different
objects. Between them sits the compiler, which must solve two problems the
author of the circuit never mentioned. The first is connectivity: a
superconducting processor couples each qubit to only a handful of neighbours,
so an entangling operation whose two operands are not neighbours cannot run at
all, and the compiler inserts SWAP gates until they are. The second is the
gate set: a device calibrates a small number of physical operations, so every
gate outside that set must be rewritten in terms of the ones inside it.

Both problems are well studied. Qubit mapping is NP-hard in general and has a
substantial literature of heuristics~\cite{siraichi2018qubit,
zulehner2019efficient, li2019sabre, murali2019noise}, of which SABRE
\cite{li2019sabre} and its recent reworking LightSABRE~\cite{zou2024lightsabre}
are the most widely deployed; gate synthesis has been understood since
Barenco~\textit{et al.}~\cite{barenco1995elementary}. What is far less
developed is the \emph{observability} of the process. A production compiler
such as Qiskit~\cite{javadiabhari2024qiskit} is invoked as a single call and
returns a single circuit. A learner, or an engineer debugging a surprising
result, is therefore left with an observation of the form ``my five-gate
circuit became seventy gates'' --- the growth a five-qubit star circuit
actually undergoes on a heavy-hex lattice in our own engine --- and no
mechanism for asking which of the two problems caused it, which SWAPs were
inserted and why those rather than others, or whether the rewriting preserved
the computation at all.

This last question is not rhetorical. Compilation is a sequence of
semantics-preserving rewrites, and a bug in any of them produces a circuit that
runs perfectly and computes the wrong thing. Equivalence checking for quantum
circuits is an active field~\cite{viamontes2007checking, amy2018towards,
burgholzer2021advanced}, but it is normally treated as an offline verification
task applied to selected cases, not as something a compiler does to every
circuit it emits --- in large part because the obvious method, comparing state
vectors, costs $2^{N}$ amplitudes in the device size and is therefore
unthinkable on a 127-qubit processor.

\subsection*{Contributions}

This paper describes \qy{}, an open-source platform of four subsystems ---
compilation, error identification, quantum machine learning, and an adaptive
learning layer over all three --- built on the position that each can be made
to account for itself at acceptable cost, and that every claim it makes should
be checked against something other than itself. Our contributions are:

\begin{enumerate}
\item \textbf{A stage-separated pipeline with cost attribution.}
  Decomposition, layout and routing, and basis translation run as distinct
  stages under an invariant: each stage may alter exactly one property of the
  circuit (gate arity, qubit placement, gate set) and must leave the others
  intact. Every gate in the final circuit is therefore attributable to a
  single cause. Within each stage, every individual compiler pass is recorded
  with its name, its duration, and the depth and size of the circuit it
  produced (Section~\ref{sec:pipeline}).

\item \textbf{A decision-recording router.} Our implementation of the SABRE
  heuristic emits, for each round of its main loop, the set of gates that
  became executable, the front layer that then blocked, the look-ahead set,
  every candidate SWAP with its front-layer cost, look-ahead cost, decay factor
  and combined score, and the winner with the reason it won. Layout synthesis
  becomes an inspectable process rather than an opaque transformation
  (Section~\ref{sec:router}, Table~\ref{tab:decision}). The dependency DAG
  behind each stage is exposed alongside it, and compiled QASM can be exported
  for several device layouts at once so their costs can be compared directly
  (Sections~\ref{sec:dag} and~\ref{sec:export}).

\item \textbf{Occupied-sub-space equivalence verification.} We observe that the
  device wires a compiled circuit never touches remain in $\ket{0}$ for the
  whole circuit and so factorise out of the state exactly. Restricting the
  state-vector comparison to the $k$ occupied wires is consequently not an
  approximation of the $2^{N}$ comparison but the same comparison with an
  identical tensor factor cancelled from both sides. This reduces
  per-compilation verification from $2^{N}$ to $2^{k}$ amplitudes and makes it
  routine rather than exceptional (Section~\ref{sec:verification}).

\item \textbf{An exactly validated error-identification module.} Three
  stabiliser codes are implemented with syndrome extraction and lookup
  decoding, and validated by enumerating \emph{every} error pattern the noise
  model can produce --- not by sampling --- and comparing the resulting
  logical error rate against closed-form algebra
  (Section~\ref{sec:qec}).

\item \textbf{A batched state-vector engine for variational training, checked
  against a general simulator.} Training evaluates one circuit per sample per
  iteration, where a general-purpose simulator's per-call overhead dominates.
  Holding the state as an $n$-dimensional array with a batch axis turns one
  training iteration into one contraction per gate rather than one pass per
  sample. We report both halves of the trade: agreement with
  \texttt{qiskit.quantum\_info} to $2.0\times10^{-15}$, and the speed-up that
  justifies writing a second simulator at all (Section~\ref{sec:qml}).

\item \textbf{A deterministic diagnostic learning layer.} Topic mastery is
  computed from weighted evidence with four published thresholds and no
  learned parameters, and a learner's misconception is \emph{measured} rather
  than inferred: a candidate mistake is reported only when mechanically
  undoing it in the submitted circuit makes that circuit meet the goal under
  simulation (Section~\ref{sec:learning}).

\item \textbf{A measured evaluation, including the cost of transparency.}
  Section~\ref{sec:evaluation} reports routing quality against three Qiskit
  routers across six topologies up to 156 qubits, verification cost against
  device size, and error-correction validation. Our router is deliberately
  unoptimised and starts from the identity layout; it therefore inserts more
  SWAPs than Qiskit's SABRE, and we quantify the gap rather than omit it.
\end{enumerate}

We make no claim to improve on the state of the art in routing quality, and we
do not present a user study: the learning layer's design is described and its
mechanism validated, but its educational effect is not measured here. Both
limitations are discussed in Section~\ref{sec:discussion}.

\section{Related Work}
\label{sec:related}

\subsection{Qubit mapping and routing}

Mapping a logical circuit onto a restricted coupling graph while minimising
inserted SWAPs is NP-hard, and the practical literature is accordingly
heuristic. Siraichi~\textit{et al.}~\cite{siraichi2018qubit} framed qubit
allocation as a compiler problem; Zulehner~\textit{et al.}
\cite{zulehner2019efficient} gave an $A^{*}$-based method for the IBM QX
devices; Murali~\textit{et al.}~\cite{murali2019noise} showed that calibration
data should inform the mapping, since qubits differ in quality.
Li and colleagues~\cite{li2019sabre} introduced SABRE, a routing algorithm
that ranks candidate SWAPs by how far they close the physical distance between
the operands of the gates waiting at the executable front layer. It uses a
look-ahead term to discount future moves and a decay factor that encourages
SWAPs to be spread around rather than focus on just one spot in the device.
That is why it became the go-to router in Qiskit. LightSABRE
\cite{zou2024lightsabre} later reworked this approach, with results reported
there showing roughly $200\times$ faster execution and nearly $19\%$ fewer
SWAPs on average. Nash~\textit{et al.}
\cite{nash2020quantum} and Maslov~\cite{maslov2017basic} treat the
complementary problem of optimising for a particular architecture's native
operations.

Our router implements the SABRE heuristic as published in
\cite{li2019sabre}; we claim no algorithmic contribution to it, and we do not
reproduce the enhancements of~\cite{zou2024lightsabre}. What we add is the
emission of the decision record. To our knowledge no production or educational
compiler exposes the candidate set with its decomposed score, which is
precisely the information needed to understand why a particular SWAP was
chosen.

\subsection{Verification of compiled circuits}

Since compilation is a chain of rewrites, its output should be checked.
Researchers like Viamontes~\textit{et al.}~\cite{viamontes2007checking} used
decision diagrams to check whether two circuits or quantum states are
equivalent. Burgholzer and Wille~\cite{burgholzer2021advanced} take a slightly
different route: they compose a circuit with the inverse of its supposed
equivalent, and if the identity comes out you have a match --- which saves
considerable work compared with building either unitary outright.
Amy~\cite{amy2018towards} and Kissinger and van de
Wetering~\cite{kissinger2019pyzx} bring other tools, path-sum and the
ZX-calculus, to the same end. All of these are stronger than ours: they verify
unitary equality rather than the action on a single input, and they handle
circuits far beyond what state-vector simulation can reach.

Our goal is more modest, but more practically grounded: we want a check fast
and reliable enough to run every single time the compiler is called, even
inside a web request. This is not to catch hardware mistakes, but to make sure
our own pipeline does not go off the rails. What makes it possible is not some
fancy algebra --- it is architectural: a typical user circuit only uses a
handful of qubits on the whole device, leaving the rest untouched. That shrinks
the problem enough to do exact state-vector comparisons on the small, active
sub-space. Our method's cost depends only on the circuit, not on the hardware
size, and it complements decision diagrams rather than competing with them.
They scale with circuit size; we stay agnostic to device size, with the
trade-off that our approach meets a ceiling in circuit \emph{width}, where the
$2^{k}$ cost catches up with us.

\subsection{Quantum error correction in teaching}

The three-qubit repetition codes and Shor's nine-qubit code
\cite{shor1995scheme, steane1996error} remain the standard introduction to
error correction, and the stabiliser formalism~\cite{gottesman1997stabilizer}
the standard language for it. Modern devices have driven interest in codes
matched to low-degree coupling graphs~\cite{chamberland2020topological}. Our
contribution here is not a code or a decoder but a validation methodology
appropriate to a teaching tool: because the codes are small, every error
pattern the noise model can produce can be enumerated exactly, so the module's
reported logical error rate can be compared against closed-form algebra rather
than against a sampled estimate of itself.

\subsection{Platforms and tooling}

Qiskit~\cite{javadiabhari2024qiskit}, t$\vert$ket$\rangle$
\cite{sivarajah2021tket} and PennyLane~\cite{bergholm2018pennylane} are the
dominant programming frameworks, and OpenQASM~\cite{cross2022openqasm} the
interchange format we accept as input. Web-based circuit editors have made it
easy for anyone to build quantum circuits, and the Qiskit textbook does a good
job of teaching the algorithms. But there is a gap right between those two:
the editors stop at drawing and simulating the circuit, and the critical step
--- compiling for real hardware, where NISQ-era~\cite{preskill2018nisq}
engineering actually gets done --- is still hidden in a single, black-box
call.

\subsection{Mastery modelling}

Our approach borrows from mastery learning~\cite{bloom1968learning} and the
original model of knowledge tracing by Corbett and Anderson
\cite{corbett1995knowledge}, where skill is treated as a probability that
updates as more evidence of what a learner knows arrives. Later work swaps
that explicit, interpretable model for a recurrent neural
network~\cite{piech2015deep}, but in doing so loses the transparency that
makes feedback meaningful. VanLehn~\cite{vanlehn2011relative} showed that
feedback at the step level --- telling a learner what, specifically, needs
fixing --- explains much of the benefit of intelligent tutoring, which for us
means zeroing in on a specific structural mistake rather than handing out
survey scores. That is why we stick with a transparent rule that weighs
evidence instead of a learned model: if our platform exists to explain itself,
every recommendation needs to come with a reason you can inspect. The
direction we have taken lines up with the needs analysis of Hughes
\textit{et al.}~\cite{hughes2022assessing}, who found that employers want
engineers who have grappled with real
hardware constraints rather than only the algorithms.

\section{System Architecture}
\label{sec:architecture}

\begin{figure}[t]
\centering
\fitcol{%
\begin{tikzpicture}[
  font=\scriptsize,
  box/.style={draw, rounded corners=1pt, align=center, inner sep=1.6pt,
              minimum height=4.4mm},
  wide/.style={box, text width=76mm},
  half/.style={box, text width=36.4mm},
  mod/.style={box, text width=16.8mm, fill=black!4},
  st/.style={box, text width=16.8mm},
  ar/.style={-{Latex[length=1.4mm]}, line width=0.35pt},
]

\node[wide] (client) at (0,0)
  {\textbf{Browser client}\\\tiny circuit editor, stage viewer, lessons, practice};
\node[wide] (api) at (0,-8.5mm)
  {\textbf{HTTP API}\;\tiny session auth, CSRF, per-route rate limits};

\node[mod] (tp)     at (-28.8mm,-17.5mm) {Compilation\\engine \tiny(100)};
\node[mod] (qec)    at ( -9.6mm,-17.5mm) {Error\\identification};
\node[mod] (learn)  at (  9.6mm,-17.5mm) {Learning\\layer};
\node[mod] (assess) at ( 28.8mm,-17.5mm) {Assessment};

\node[wide, fill=black!2] (engine) at (0,-27mm)
  {\textbf{Inside the compilation engine (100)}};

\node[st] (s1) at (-28.8mm,-34mm) {ingest\\\tiny(110)};
\node[st] (s2) at ( -9.6mm,-34mm) {decompose\\\tiny(120)};
\node[st] (s3) at (  9.6mm,-34mm) {layout +\\route \tiny(130)};
\node[st] (s4) at ( 28.8mm,-34mm) {basis\\\tiny(140)};

\node[half] (ver)   at (-19.2mm,-42mm) {verification \tiny(150)};
\node[half] (trace) at ( 19.2mm,-42mm) {decision + stage record \tiny(160)};

\node[wide] (db) at (0,-50.5mm)
  {\textbf{Relational store}\;\tiny profiles, lesson and attempt history\\
   \tiny\itshape no circuit is stored server-side};

\draw[ar] (client) -- (api);
\foreach \t in {tp,qec,learn,assess} \draw[ar] (api.south) -- (\t.north);
\draw[ar] (tp.south) -- (tp.south |- engine.north);

\draw[ar] (engine.south -| s1) -- (s1.north);
\foreach \a/\b in {s1/s2, s2/s3, s3/s4} \draw[ar] (\a) -- (\b);

\draw[ar] (s4.south) -- ++(0,-2mm) -| (ver.north);
\draw[ar] (ver.east) -- (trace.west);
\draw[ar] (trace.south) -- (trace.south |- db.north);

\draw[ar] (learn.south) -- (9.6mm,-22.2mm) -- (43mm,-22.2mm)
          -- (43mm,-50.5mm) -- (db.east);

\end{tikzpicture}}
\caption{Platform architecture. The compilation engine (100) is a pure
function of its request: it holds no state, stores no circuit, and has no
import path to the hardware-submission module, so a compilation can never
become a hardware job by accident.}
\label{fig:architecture}
\end{figure}
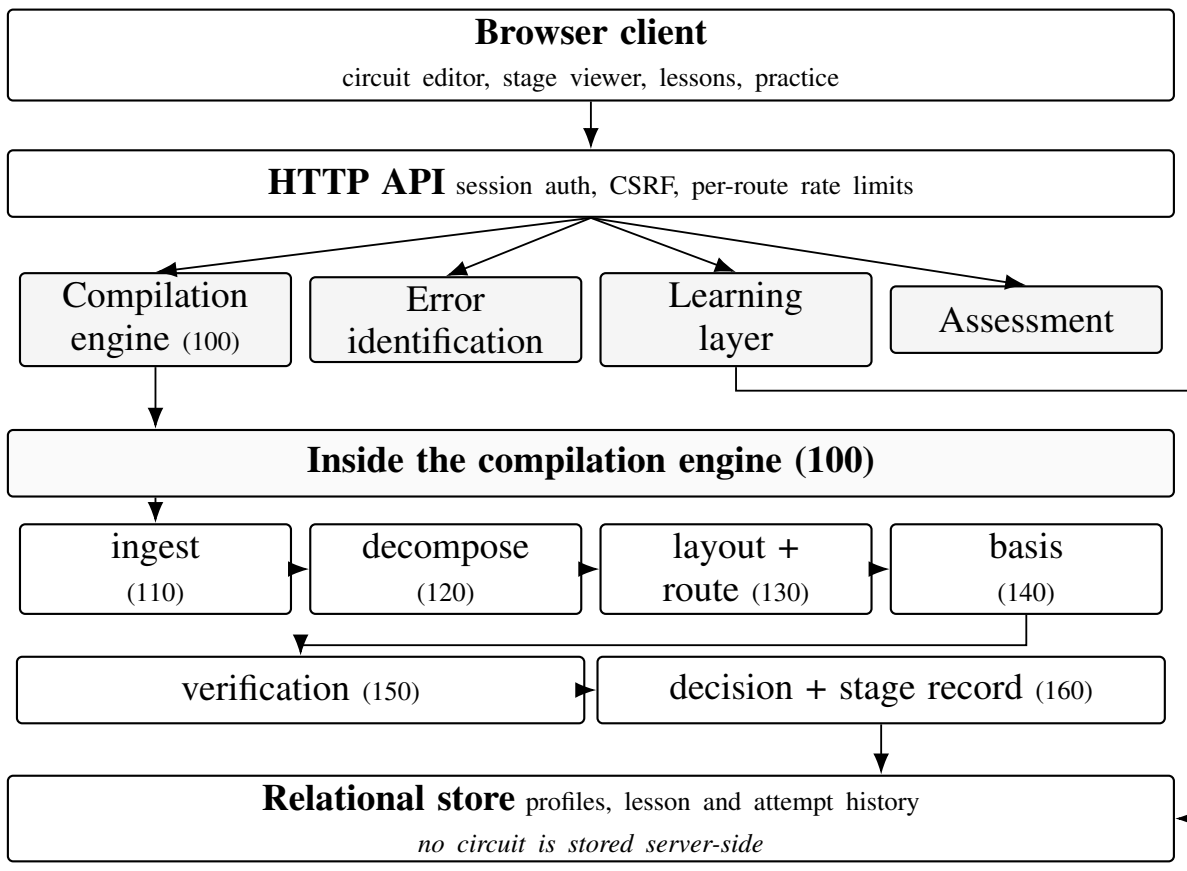

\qy{} is a web application: a browser client and a Python service layer
over a relational store (Fig.~\ref{fig:architecture}). The service layer is
divided into six packages that do not import one another except where stated:
the compilation engine, the machine-learning studio, the error-identification
module, the assessment module, the learning layer (which deliberately depends
on assessment), and authentication. Table~\ref{tab:code} gives their sizes.

Three structural properties are worth stating because they are enforced rather
than merely intended.

\subsubsection*{The compilation engine is stateless and isolated}
The engine (numeral 100 in Fig.~\ref{fig:architecture}) is a pure function from
a request to a response. It persists nothing, so no user circuit is stored
server-side, and it has no import path to the hardware-submission module. The
latter is checked by a test that walks the engine's transitive imports and
fails if the hardware module appears among them. Running a circuit on real
hardware is a separate, separately authenticated feature; the consequence of
the isolation is that a compilation can never become a billable hardware job
through a coding error.

\subsubsection*{Every limit is explicit}
A request is bounded before work begins: at most $400$ source instructions, at
most $60$ per-pass circuit snapshots per stage, no snapshot of a circuit
exceeding $900$ gates, and equivalence verification only up to $12$ occupied
qubits. \qy{} does not treat these as fuzzy settings --- they are hard
promises. Either the engine answers within the time a web request allows, or
it declines and says exactly why. No silent timeouts, and no degraded result
passed off as a complete one.

\subsubsection*{Scoring is server-side}
Quiz answers, challenge grading and progress all live in the service layer.
Nothing that determines a result is present in the browser before it is due,
and the dashboard cannot disagree with the stored record because it is derived
from it on each request.

\begin{table}[t]
\caption{Implementation size, by package. Lines include comments and blank
lines as the source contains them.}
\label{tab:code}
\centering
\footnotesize
\begin{tabular}{@{}llrr@{}}
\toprule
Layer & Package & Files & Lines \\
\midrule
\multirow{6}{*}{Service}
  & compilation engine            & 10 & 4\,187 \\
  & machine learning studio       & 12 & 5\,777 \\
  & error identification          & 5  & 1\,277 \\
  & learning layer                & 10 & 2\,326 \\
  & assessment                    & 7  & 1\,811 \\
  & authentication                & 27 & 2\,029 \\
\midrule
Tests & automated test suite      & 22 & 6\,391 \\
\midrule
Client & browser application      & 125 & 24\,696 \\
\bottomrule
\end{tabular}
\end{table}

\subsection{Device models}
\label{sec:devices}

The engine ships the coupling graphs listed in Table~\ref{tab:topologies}.
Linear, ring, grid and all-to-all graphs are generated analytically; the
heavy-hex lattices follow the degree-$\le 3$ construction IBM adopted for its
superconducting processors~\cite{chamberland2020topological}, and the two
named layouts reproduce the qubit and coupling counts of specific device
generations. A user-supplied edge list is also accepted, subject to a
connectivity check --- a disconnected graph makes routing impossible, and
rejecting it with that explanation is more useful than failing to converge.

Each device's connectivity graph is tracked in two forms, and keeping them
separate is not just pedantry. When routing, we work with the
\emph{undirected} map, because a SWAP moves two qubits together regardless of
entangling direction. For basis translation and final verification, direction
matters: the calibrated direction of a gate signals whether extra fix-ups are
needed, so those stages take the \emph{directed} map. Mixing the two would
either create unnecessary SWAPs or miss a direction violation entirely.
Keeping them apart means the final check can report wrong-direction gates as a
separate, non-fatal warning.

\begin{table}[t]
\caption{Device coupling graphs available to the engine.}
\label{tab:topologies}
\centering
\footnotesize
\begin{tabular}{@{}llr@{}}
\toprule
Graph & Structure & Qubits \\
\midrule
Linear chain   & path                                   & $3$--$20$ \\
Ring           & cycle                                  & $3$--$20$ \\
2-D grid       & nearest-neighbour lattice              & $3$--$20$ \\
Heavy-hex      & degree-$\le 3$ lattice, $d=3$          & $19$ \\
All-to-all     & complete graph (trapped-ion-like)      & $3$--$20$ \\
IBM Eagle      & heavy-hex device layout                & $127$ \\
IBM Heron r2   & heavy-hex device layout                & $156$ \\
Custom         & user edge list, connectivity-checked   & $\le 20$ \\
\bottomrule
\end{tabular}
\end{table}

\subsection{Target gate sets}

Five basis sets are supported: IBM's native
$\{R_{Z}, \sqrt{X}, X, \mathrm{CX}\}$; the Heron set with echoed
cross-resonance~\cite{rigetti2010fully}; the textbook
$\{U, \mathrm{CX}\}$; $\{R_{X}, R_{Z}, \mathrm{CZ}\}$; and the trapped-ion set
built on M\o{}lmer--S\o{}rensen gates~\cite{molmer1999multiparticle}.
Offering several is pedagogically load-bearing: compiling one circuit to both
the CX and the CZ set shows directly that the two-qubit count is a property of
the device's calibrated entangler and not of the algorithm, and pairing the
trapped-ion set with the all-to-all graph shows that the SWAP overhead which
dominates superconducting compilation is an artefact of connectivity rather
than a law of quantum computing.

\section{The Stage-Separated Pipeline}
\label{sec:pipeline}

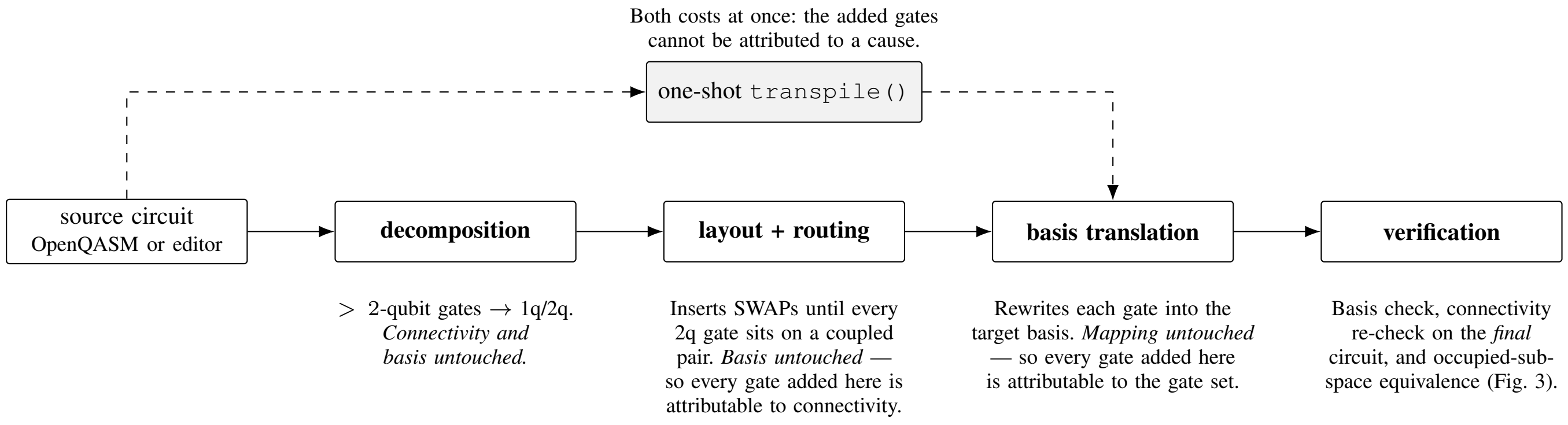
\begin{figure*}[t]
\centering
\fitpage{%
\begin{tikzpicture}[
  font=\footnotesize,
  st/.style={draw, rounded corners=1pt, align=center, inner sep=2.6pt,
             minimum height=7mm, text width=26mm},
  note/.style={align=center, font=\scriptsize, text width=34mm, inner sep=1pt},
  ar/.style={-{Latex[length=1.8mm]}, line width=0.4pt},
]

\def\dx{38mm}

\node[st] (src) at (0,0)      {source circuit\\\scriptsize OpenQASM or editor};
\node[st] (dec) at (\dx,0)    {\textbf{decomposition}};
\node[st] (rt)  at (2*\dx,0)  {\textbf{layout + routing}};
\node[st] (bs)  at (3*\dx,0)  {\textbf{basis translation}};
\node[st] (vf)  at (4*\dx,0)  {\textbf{verification}};

\draw[ar] (src) -- (dec);
\draw[ar] (dec) -- (rt);
\draw[ar] (rt)  -- (bs);
\draw[ar] (bs)  -- (vf);

\node[note, below=4mm of dec] {$>2$-qubit gates $\rightarrow$ 1q/2q.\\
  \emph{Connectivity and basis untouched.}};
\node[note, below=4mm of rt] {Inserts SWAPs until every 2q gate sits on a
  coupled pair. \emph{Basis untouched} --- so every gate added here is
  attributable to connectivity.};
\node[note, below=4mm of bs] {Rewrites each gate into the target basis.
  \emph{Mapping untouched} --- so every gate added here is attributable to
  the gate set.};
\node[note, below=4mm of vf] {Basis check, connectivity re-check on the
  \emph{final} circuit, and occupied-sub-space equivalence
  (Fig.~\ref{fig:subspace}).};

\node[st, fill=black!5, above=9mm of rt, text width=30mm] (one)
  {one-shot \texttt{transpile()}};
\draw[ar, dashed] (src.north) |- (one.west);
\draw[ar, dashed] (one.east) -| (bs.north);
\node[note, above=1.2mm of one, text width=56mm]
  {Both costs at once: the added gates cannot be attributed to a cause.};

\end{tikzpicture}}
\caption{Stage separation. Each stage is permitted to change exactly one
property of the circuit, which is what makes the gate count it adds
attributable. The dashed path is the conventional single call, shown for
comparison.}
\label{fig:pipeline}
\end{figure*}

\subsection{Stage separation and the attribution invariant}

Let $C_{0}$ be the source circuit and $C_{4}$ the circuit finally emitted. A
conventional compiler computes $C_{4} = T(C_{0})$ in one call. \qy{} computes

\begin{equation}
C_{0} \xrightarrow{\;D\;} C_{1} \xrightarrow{\;R\;} C_{2}
       \xrightarrow{\;B\;} C_{3} \xrightarrow{\;V\;} C_{4},
\label{eq:stages}
\end{equation}

where, in \eqref{eq:stages}, $D$ decomposes gates of arity greater than two, $R$ performs layout and
routing, $B$ performs basis translation, and $V$ verifies and changes nothing.
Each stage carries an invariant (Fig.~\ref{fig:pipeline}):

\begin{itemize}
\item $D$ may reduce gate arity. It may not change the qubit mapping and may
  not change the gate set beyond what the decomposition of a wide gate
  requires.
\item $R$ may insert SWAP gates and permute the logical-to-physical mapping.
  \emph{It may not alter the gate set.}
\item $B$ may rewrite any gate into the target basis. \emph{It may not alter
  the qubit mapping.}
\item $V$ may not alter the circuit at all.
\end{itemize}

The consequence is attribution. Because $R$ cannot change the basis, every
gate present in $C_{2}$ but not in $C_{1}$ exists because of connectivity.
Because $B$ cannot change the mapping, every gate present in $C_{3}$ but not in
$C_{2}$ exists because of the gate set. The two costs that a single
\texttt{transpile()} call returns summed and indistinguishable are therefore
separated, and the familiar observation that a small circuit became a large one
resolves into two numbers with two different remedies: a connectivity cost,
addressed by a better layout, and a translation cost, addressed by choosing
a different target device.

At each stage we also write down every compiler pass: what it was, how long it
took, the depth and size of the circuit it passed forward, plus a
human-friendly summary of what it did. Instead of a black box, you get a
concrete, measurable list of steps --- which is the granularity at which the
growth of a circuit can actually be understood.

\subsection{The decision-recording router}
\label{sec:router}

Routing is where the interesting decisions are made, and where conventional
tooling is most opaque. Our implementation follows the SABRE heuristic of
Li~\textit{et al.}~\cite{li2019sabre}. At each round the router maintains a
mapping $\pi$ from logical to physical qubits, computes the \emph{front layer}
$F$ of instructions all of whose predecessors have executed, and executes
every member of $F$ whose operands are adjacent under $\pi$, repeating until no
further progress is possible. If instructions remain, the front layer is
blocked, and the router must insert a SWAP. It forms the candidate set

\begin{equation}
S = \{\,(p, p') : p' \in \mathcal{N}(p),\;
      p \in \pi(\mathrm{supp}(F))\,\},
\end{equation}

that is, every device edge incident on a physical qubit that some blocked
instruction is waiting on, and scores each candidate by

\begin{equation}
H(s) = \bigl[\, \bar{d}_{\pi_{s}}(F) + w\,\bar{d}_{\pi_{s}}(E) \,\bigr]
       \cdot \max\bigl(\delta_{p}, \delta_{p'}\bigr),
\label{eq:heuristic}
\end{equation}

where $\pi_{s}$ is the mapping after applying $s$, $\bar{d}_{\pi}(\cdot)$ is
the mean device distance between the operands of the two-qubit gates in a set
under $\pi$, $E$ is a look-ahead set of up to $L$ two-qubit gates immediately
behind the front layer, $w$ weights the look-ahead term, and $\delta$ is a
per-qubit decay that rises by $\Delta$ each round and resets to $1$ for the two
qubits just swapped. We use the published defaults $L=5$, $w=0.5$,
$\Delta=10^{-3}$. The candidate minimising $H$ in \eqref{eq:heuristic} is applied; ties are
broken on
the qubit pair, which makes a run reproducible.

Two implementation choices depart from the reference formulation, and neither
touches the algorithm. First, we compute neighbour sets, edge membership and
all-pairs shortest paths with breadth-first search over an adjacency list
rather than through a graph library, which removes a dependency from the
service without changing any distance. Second, and this is the point of the
exercise, the information a reference implementation would print for debugging
is instead collected into a structured record: for every round, the
instructions that became executable, the blocked front layer with the device
distance that blocks each one, the look-ahead set, the full candidate set with
$\bar{d}(F)$, $\bar{d}(E)$, $\delta$ and $H$ reported separately for each
candidate, the selected SWAP with its score, and the mapping before and after
(Table~\ref{tab:decision}).

\begin{table}[t]
\caption{The decision record for one routing round, in the form the engine
returns it (illustrative values, $w=0.5$, look-ahead depth $5$, decay step
$0.001$). The front layer is blocked because $q_{0}$ and $q_{2}$ sit two hops
apart. Every candidate is reported with its score decomposed, not just the
winner.}
\label{tab:decision}
\centering
\setlength{\tabcolsep}{4pt}
\footnotesize
\begin{tabular}{@{}lrrrr@{}}
\toprule
Candidate SWAP & $\bar{d}(F)$ & $\bar{d}(E)$ & $\delta$ & $H$ \\
\midrule
$(p_{0},p_{1})$ & 1.00 & 1.50 & 1.000 & \textbf{1.750} \\
$(p_{1},p_{2})$ & 1.00 & 2.00 & 1.000 & 2.000 \\
$(p_{2},p_{3})$ & 2.00 & 1.50 & 1.001 & 2.753 \\
$(p_{0},p_{5})$ & 3.00 & 2.00 & 1.000 & 4.000 \\
\bottomrule
\end{tabular}
\\[2pt]
\footnotesize
Selected $(p_{0},p_{1})$ at $H=1.750$; ties break on the qubit pair, so the
run is reproducible. The decay then resets:
$\delta_{p_{0}}, \delta_{p_{1}} \leftarrow 1$, all others $+\,0.001$.
\end{table}

The design decision embedded in this record is that the \emph{losing}
candidates are reported. A router that returns only the chosen SWAP cannot be
checked; a router that returns the whole scored candidate set can be checked by
hand, and its heuristic can be criticised on a concrete case. This is also
what makes the router usable as a teaching object: a reader can see that the
look-ahead term and the decay term occasionally disagree with the front-layer
term, and can see which one won.

An unavoidable corollary is cost. The record for one round is $O(|S|)$ and
$|S|$ grows with device degree, so on a 156-qubit lattice a long routing
problem produces a large record. Section~\ref{sec:cost} measures the record
size and the resulting runtime, and Section~\ref{sec:quality} measures the
routing quality this implementation achieves given that it starts from the
identity layout and performs no optimisation.

\subsection{The dependency graph}
\label{sec:dag}

Every circuit starts as a diagram --- just the order you wrote the gates in.
The DAG reveals what that order really depends on. Wherever one gate is
immediately followed by another on a shared wire, you draw an arrow from the
first to the second. If two gates are unconnected, they can run in any order
or even at the same time. The number of layers, where each node's layer is one
past the deepest node feeding into it, gives you the true circuit depth. We
expose this at every stage, so a circuit can be read as a list of gates or as
the underlying set of constraints, and so the effect of routing on the
critical path is visible rather than inferred.

Two things make a circuit DAG simpler to lay out than a regular graph.
Horizontally there are no choices: each gate's layer is forced, and every
dependency arrow points forward. Vertically it is the qubit index that
decides. Gates in the same layer always act on disjoint wires, since two gates
sharing a qubit must be ordered and so cannot share a layer. Disjoint sets of
wires have distinct minima, so you can plot every gate at the lane of its
lowest qubit with no collisions. The upshot is that the drawing reads like the
actual circuit: a gate travels along its wire, and a horizontal edge is a
dependency between operations on the same qubit.

The graph is returned as data, with the per-node table of gate, layer, wires,
parameters, predecessors and successors available as JSON or CSV, and the
graph itself as Graphviz DOT.

\subsection{Exporting for a chosen device}
\label{sec:export}

A QASM file only means something when you know what hardware it is tied to.
Exporting in \qy{} takes a \emph{set} of target devices: each gets its own
compilation and its own file, and the SWAP counts sit side by side for
comparison. A four-qubit star circuit, for instance, needs zero SWAPs if
all-to-all connectivity is allowed, two on a linear chain and five on a
heavy-hex device --- all with the same circuit and the same gate basis
(Section~\ref{sec:quality}).

Every output file comes with a header, in the QASM comments, showing its
provenance: which device, the coupling count, the basis set, the layout and
routing method, SWAPs inserted, depth and gate counts before and after, and
the result of the equivalence check of Section~\ref{sec:verification}. If a
target device is too small, that is called out for just that target rather
than as a failure of the whole request, so one impossible layout does not cost
the user the others.

\section{Occupied-Sub-Space Equivalence Verification}
\label{sec:verification}

\begin{figure}[t]
\centering
\fitcol{%
\begin{tikzpicture}[
  font=\scriptsize,
  wire/.style={line width=0.35pt, black!55},
  live/.style={line width=0.8pt, black},
  g/.style={draw, fill=white, minimum width=3mm, minimum height=3mm,
            inner sep=0pt, font=\tiny},
  brace/.style={decorate, decoration={brace, amplitude=2.5pt},
                line width=0.35pt},
  lbl/.style={font=\tiny, align=left},
]

\def\devlen{28mm}

\foreach \i/\y in {0/0, 1/-3.2mm, 2/-6.4mm, 3/-9.6mm} {
  \draw[live] (0,\y) -- (\devlen,\y);
  \node[left=0.6mm, font=\tiny] at (0,\y) {$p_{\i}$};
}

\foreach \y in {-14.2mm, -17.4mm, -22.6mm} {
  \draw[wire] (0,\y) -- (\devlen,\y);
}
\node[left=0.6mm, font=\tiny] at (0,-14.2mm) {$p_{4}$};
\node[left=0.6mm, font=\tiny] at (0,-17.4mm) {$p_{5}$};
\node[left=0.6mm, font=\tiny] at (0,-22.6mm) {$p_{N\text{-}1}$};
\node at (14mm,-20mm) {$\vdots$};

\node[lbl, anchor=west] at (29.5mm,-17.4mm) {$\lvert 0\rangle$ throughout};

\node[g] at (4mm,0) {H};
\draw[live] (9mm,0) -- (9mm,-3.2mm);
\filldraw (9mm,0) circle (0.45mm);
\node[g] at (9mm,-3.2mm) {$\oplus$};
\node[g] at (14mm,-6.4mm) {H};
\draw[live] (19mm,-3.2mm) -- (19mm,-6.4mm);
\filldraw (19mm,-3.2mm) circle (0.45mm);
\node[g] at (19mm,-6.4mm) {$\oplus$};
\draw[live] (24mm,-6.4mm) -- (24mm,-9.6mm);
\filldraw (24mm,-6.4mm) circle (0.45mm);
\node[g] at (24mm,-9.6mm) {$\oplus$};

\draw[brace] (-5mm,0.8mm) -- (-5mm,-10.4mm)
  node[midway, left=2pt, lbl] {$k$};
\draw[brace] (-9.5mm,1.1mm) -- (-9.5mm,-23.4mm)
  node[midway, left=2pt, lbl] {$N$};

\node[draw, rounded corners=1pt, align=center, inner sep=2.4pt,
      text width=31mm, fill=black!3, font=\scriptsize]
      (full) at (3mm,-32mm)
  {unrestricted: $2^{N}$\\[1pt]
   \tiny $N=127 \Rightarrow 2.7\!\times\!10^{39}$ bytes};
\node[draw, rounded corners=1pt, align=center, inner sep=2.4pt,
      text width=31mm, font=\scriptsize]
      (red) at (41mm,-32mm)
  {restricted: $2^{k}$\\[1pt]
   \tiny $k=5 \Rightarrow 512$ bytes};

\draw[-{Latex[length=1.4mm]}, line width=0.35pt] (full) -- (red)
  node[midway, above, font=\tiny] {exact};

\end{tikzpicture}}
\caption{Occupied-sub-space equivalence verification. Every wire the compiled
circuit never touches provably remains in $\ket{0}$ and therefore factorises
out of the state exactly, so comparing only the $k$ occupied wires is not an
approximation of the $2^{N}$ comparison --- it is the same comparison, with a
tensor factor of $\ket{0}^{\otimes(N-k)}$ cancelled from both sides.}
\label{fig:subspace}
\end{figure}
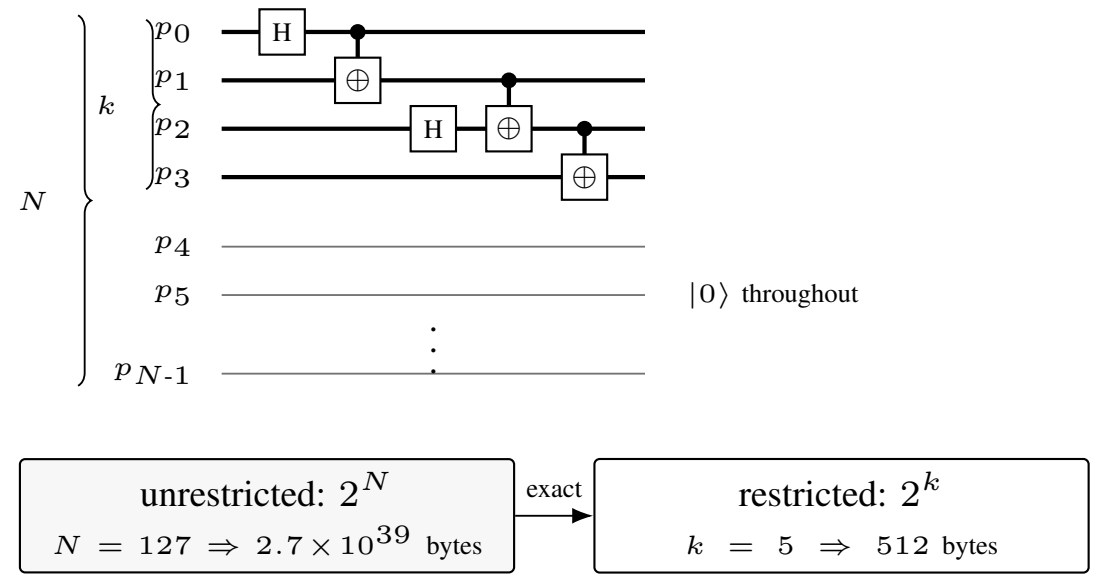

Each compilation is checked three ways before it is returned. Two are simple
bookkeeping: first, making sure every instruction in the final circuit matches
what the target gate set allows; second, re-checking connectivity, but only
after all routing and transformation is done, to be sure basis translation did
not introduce new problems. The third check goes deeper --- it proves the
circuit still computes the same thing it was meant to.

\subsection{The reduction}

Say the hardware has $N$ qubits but the circuit only uses $k$ of them --- the
wires carrying at least one instruction, together with the wires the user's
logical qubits finally sit on. Collect the indices of those active wires in
$\mathcal{K} \subset \{0,\dots,N-1\}$, with $|\mathcal{K}| = k$, and let
$\mathcal{I}$ be the rest.

There are no instructions at all on the wires in $\mathcal{I}$. That is not a
guess; it is checked directly from the emitted instruction list. We always
start from $\ket{0}^{\otimes N}$, so every idle wire in $\mathcal{I}$ stays at
$\ket{0}$ the whole way, and the final state factorises cleanly:

\begin{equation}
U_{C_{4}} \ket{0}^{\otimes N}
  = \Bigl( U'_{C_{4}} \ket{0}^{\otimes k} \Bigr)
    \otimes \ket{0}^{\otimes (N-k)} ,
\label{eq:factorise}
\end{equation}

where $U'_{C_{4}}$ is the circuit re-expressed on the $k$ occupied wires alone,
obtained by the order-preserving relabelling $\mathcal{K} \to \{0,\dots,k-1\}$.

Let $\pi$ be the final logical-to-physical mapping reported by the routing
stage, and define the induced basis-state map on indices
\begin{equation}
\Phi(i) \;=\; \sum_{v=0}^{n-1}
  \bigl[(i \gg v)\,\&\,1\bigr] \cdot 2^{\,\mathrm{pos}(\pi(v))},
\label{eq:phi}
\end{equation}
where $n$ is the number of logical qubits and $\mathrm{pos}(\cdot)$ gives a
physical qubit's position in the sorted occupied set. Because the idle wires
stay in $\ket{0}$, every logical basis state maps to exactly one physical basis
state and $\Phi$ is injective. The expected state defined by \eqref{eq:phi} is then
$\ket{\psi_{\mathrm{exp}}} = \sum_{i} \alpha_{i} \ket{\Phi(i)}$ for
$\ket{\psi_{\mathrm{src}}} = \sum_{i} \alpha_{i} \ket{i}$ the source circuit's
state, and the check is

\begin{equation}
\mathcal{F} \;=\;
  \bigl| \langle \psi_{\mathrm{exp}} \,|\, U'_{C_{4}} \ket{0}^{\otimes k}
  \bigr|^{2}
  \;\overset{?}{>}\; 1 - 10^{-8}.
\label{eq:fidelity}
\end{equation}

Equation~\eqref{eq:factorise} is what makes this exact. Restricting to
$\mathcal{K}$ cancels an identical tensor factor $\ket{0}^{\otimes(N-k)}$ from
both sides of the comparison; no amplitude is discarded, truncated or
approximated. The restricted check is the unrestricted check, performed after
an exact simplification. The practical effect is that verification cost depends
on the circuit the user wrote and not at all on the device they selected ---
which is the difference between a check that can run on every compilation and
one that cannot run on a 127-qubit device at all.

\subsection{Scope and honest limits}

We only attempt verification when the conditions fit, and when they do not the
engine says so clearly --- there is no silent skipping. It is skipped if the
circuit measures or resets, since it is then not a single unitary; if layout
information is missing; and if $k$ exceeds $12$, because $2^{13}$ amplitudes
is simply too big for our latency target. The tolerance $10^{-8}$
in~\eqref{eq:fidelity} is on the loose side for double precision, chosen so
that accumulated error in a long gate sequence does not light up a false
alarm.

There are three real limits. First, we only check whether the output starting
from $\ket{0}^{\otimes N}$ matches --- not unitary equality. A faulty rewrite
that acts identically on $\ket{0}^{\otimes N}$ will slip by, and tools like
decision diagrams~\cite{burgholzer2021advanced} and
path-sums~\cite{amy2018towards} are genuinely stronger here. Second, the
$k \le 12$ ceiling is real: device size does not matter, but a circuit using
more than twelve wires goes unverified. Third, this only checks our own
compiler pipeline; it says nothing about the behaviour of the physical device.
The upside is that you get an unconditional check every time, right inside the
compile, so if something is wrong you find out immediately rather than in an
experimental result. Section~\ref{sec:cost} measures what that costs.

\section{Error Identification}
\label{sec:qec}

The platform includes a module for the idea that most reliably defeats
newcomers: that an error on a qubit can be \emph{located} without the qubit
being measured. We have built support for three codes in the stabiliser
formalism~\cite{gottesman1997stabilizer, nielsen2010quantum}: the basic
three-qubit bit-flip and phase-flip repetition codes, and Shor's nine-qubit
code~\cite{shor1995scheme, steane1996error}. For each code you get the check
operators, the syndrome-to-correction lookup table, and clear logical
$\bar{X}$ and $\bar{Z}$ actions. Table~\ref{tab:qec} gives their parameters.

\subsection{Representation}

A Pauli error on $n$ data qubits is held as a pair of binary vectors
$(\mathbf{x}, \mathbf{z}) \in \mathbb{F}_{2}^{n} \times \mathbb{F}_{2}^{n}$,
with $X \mapsto (1,0)$, $Z \mapsto (0,1)$, $Y \mapsto (1,1)$. For a check
operator with support vectors $(\mathbf{a}, \mathbf{b})$ the syndrome bit is
the symplectic product
\begin{equation}
\sigma = \mathbf{a}\!\cdot\!\mathbf{z} \;\oplus\;
         \mathbf{b}\!\cdot\!\mathbf{x} \pmod 2 ,
\label{eq:syndrome}
\end{equation}
Stringing these together for a batch of errors, by \eqref{eq:syndrome}, is
just multiplying a matrix by a vector over $\mathbb{F}_{2}$, so a whole batch
of patterns costs one matrix product. Decoding is a table lookup from syndrome
to correction.

\subsection{Classifying what is left over}

It is not enough to say only ``corrected'' or ``not corrected''. What really
matters is what happens after correction. Take the residual
$\mathbf{r} = \mathbf{e} \oplus \mathbf{f}$, where $\mathbf{e}$ is the
error and $\mathbf{f}$ the correction, and classify it by how it commutes
with the logical operators:

\begin{itemize}
\item If $\mathbf{r}$ is zero, the error was truly \emph{corrected}.
\item If $\mathbf{r}$ is non-zero but still commutes with both $\bar{X}$ and
  $\bar{Z}$, you are \emph{corrected up to a stabiliser}. The final state is
  fine even though the correction was not the literal inverse of the error.
  This is the first point where the subject actually feels quantum rather than
  binary.
\item If $\mathbf{r}$ anticommutes with either $\bar{X}$ or $\bar{Z}$ and the
  syndrome was non-zero, you \emph{mis-corrected}: the decoder acted, and made
  things worse.
\item If it anticommutes and the syndrome is zero, the error went
  \emph{undetected} --- it is itself a logical operator, and the code is blind
  to it.
\end{itemize}

The last two categories are the honest content of the subject: a code does not
remove errors, it trades a high rate of correctable errors for a low rate of
uncorrectable ones, and a module that reported only success would teach the
opposite.

\subsection{Exact validation}

We do not estimate logical error rates by hoping repeated samples will cover
the space. For an independent single-qubit noise model with per-qubit Pauli
probabilities, the module enumerates every single possibility --- for the
nine-qubit code that is $4^{9} = 262\,144$ distinct patterns. For each one it
calculates the exact probability, syndrome, correction and outcome class using
the vectorised algebra above. Add them up and you have the exact rate, not an
estimate. This is not just for completeness: it serves as a gold-standard
check. With matched noise, the three-qubit code must fail exactly when two or
three of its qubits are hit, matching the closed form
\begin{equation}
P_{\mathrm{logical}}(p) = 3p^{2}(1-p) + p^{3} = 3p^{2} - 2p^{3}.
\label{eq:closedform}
\end{equation}
Section~\ref{sec:qecresults} reports the measured agreement with
\eqref{eq:closedform}.

\subsection{Circuit construction}

Alongside the algebra the module emits the actual Qiskit circuit, annotated
with stage bands (\textsc{prepare}, \textsc{encode}, \textsc{error},
\textsc{syndrome}, \textsc{measure}) so the same drawing can be read stage by
stage. Two simulation modes are offered. In \emph{injection} mode the user
places specific errors and the circuit is run as a state vector, which is
deterministic and so lets the syndrome be read as a fact rather than a
statistic. In \emph{noise} mode errors occur with probability $p$ and the
circuit is sampled; here the matrix-product-state simulator is used instead,
which on the 17-qubit Shor circuit is roughly an order of magnitude faster
than dense state-vector simulation and sufficient for the low-entanglement
states these circuits produce. Logical-basis states of the phase-flip code are
rendered in the $\{\ket{+},\ket{-}\}$ basis, where
$\alpha\ket{{+}{+}{+}} + \beta\ket{{-}{-}{-}}$ is two terms rather than the
eight it becomes in the computational basis.

\begin{table}[t]
\caption{Implemented codes. Total qubits include the ancillae used for
syndrome extraction.}
\label{tab:qec}
\centering
\footnotesize
\begin{tabular}{@{}lrrrr@{}}
\toprule
Code & Data & Checks & Total & Error patterns \\
\midrule
Bit-flip repetition    & 3 & 2 & 5  & $4^{3} = 64$ \\
Phase-flip repetition  & 3 & 2 & 5  & $4^{3} = 64$ \\
Shor nine-qubit        & 9 & 8 & 17 & $4^{9} = 262\,144$ \\
\bottomrule
\end{tabular}
\end{table}

\section{Quantum Machine Learning Studio}
\label{sec:qml}

The fourth subsystem trains variational models. It is included here not because
variational classification is novel --- it is not --- but because two of its
engineering decisions follow the same rule as the rest of the platform: do the
arithmetic in a way that can be checked, and check it.

\subsection{What it offers}

Three model families are implemented: a variational quantum classifier, a
hybrid quantum--classical network, and a quantum-kernel support vector
classifier. Each is assembled from a feature map, an ansatz and an optimiser,
listed in Table~\ref{tab:qmlparts}. Multi-class problems are reduced
one-against-one or one-against-rest. Everything is bounded before work starts,
as elsewhere in the platform: the kernel classifier refuses more than 200
samples, because a quantum kernel is a dense $N \times N$ matrix of circuit
evaluations and the request would not return.

\begin{table}[t]
\caption{What a model is assembled from. Every combination is reachable from
the browser without writing code.}
\label{tab:qmlparts}
\centering
\setlength{\tabcolsep}{4pt}
\footnotesize
\begin{tabular}{@{}ll@{}}
\toprule
Component & Choices \\
\midrule
Model & VQC, hybrid QNN, quantum-kernel SVC \\
Feature map & angle, angle-$X$, dense angle, $Z$, $ZZ$, amplitude \\
Ansatz & hardware-efficient, real-amplitudes, \\
       & efficient-SU(2), product, user-supplied \\
Optimiser & Adam (parameter-shift), SPSA, COBYLA \\
Multi-class & one-vs-one, one-vs-rest \\
Datasets & moons, circles, blobs, XOR, iris, wine, digits \\
Execution & simulator, or IBM Quantum hardware \\
\bottomrule
\end{tabular}
\end{table}

\subsection{Gradients that differentiate the right thing}

Adam uses the parameter-shift rule, which gives an exact derivative of a
circuit's \emph{expectation value} with respect to a rotation angle. This
needs saying because it is a really easy mistake: the parameter-shift rule
does not apply to the loss function directly. Loss depends on the expectation
value, so you must use the chain rule --- take the shift-rule gradient of the
expectation value first, then multiply by the derivative of the loss with
respect to that expectation. Apply the shift rule straight to the loss and it
spits out a number that looks reasonable but is not a real gradient. Training
might still look like it is working, which is why this mistake slips through
into real code.

SPSA and COBYLA are offered alongside it because the shift rule costs two
circuit evaluations per parameter per step, which is the dominant cost of
training and the reason the engine below exists.

\subsection{A second simulator, and why}

Training does not call Qiskit. The loop evaluates a variational circuit once
per sample per iteration, so a few hundred samples over a hundred iterations is
tens of thousands of evaluations, and at the circuit widths a browser session
uses --- four to twelve qubits --- the per-call overhead of a general-purpose
simulator dominates the arithmetic it performs.

The engine in \texttt{statevector.py} is written for that case. It holds the
state as an $n$-dimensional array of shape $(2,2,\dots,2)$ rather than a flat
vector of length $2^{n}$, which turns applying a single-qubit gate into a
tensor contraction along one axis and a two-qubit gate into a contraction along
two. More importantly it carries a batch axis, so one circuit can be applied to
every sample in a batch with one contraction per gate instead of one pass per
sample. Since each sample in a batch differs only in its feature map and shares
the entire ansatz, that is the whole of the hot path.

Writing a second simulator is only defensible if it gives the same answers as
the first, so both properties are measured rather than asserted.
Section~\ref{sec:qmlresults} reports agreement with
\texttt{qiskit.quantum\_info} to a worst infidelity of
$2.0 \times 10^{-15}$, and the speed-up that pays for the duplication: modest
on a single circuit, and between roughly $25\times$ and $50\times$ on the batched path
that training actually runs.

\subsection{Hardware execution, and the credentials it needs}

A trained model can be re-evaluated on IBM Quantum hardware. This is the one
place in the platform that holds a user credential, and it is deliberately the
one place: the compilation engine of Section~\ref{sec:pipeline} has no import
path to it, enforced by a test (Section~\ref{sec:architecture}), so no amount
of activity in the transpiler can turn into a hardware job.

The token is never returned by any endpoint, never written to a log, and never
sent back to the browser that supplied it. It is encrypted at rest with a
symmetric key the deployment supplies, scoped to the account that stored it,
and revocable. Submitting jobs to real hardware uses a separate rate limit
from simulation, and before anything is sent out the system checks the
configuration against the chosen device --- number of qubits, maximum depth.
If there is no way the request could have succeeded, it is blocked up front,
sparing both the hardware quota and the user's patience. These are not
incidental; a teaching platform that leaks a
hardware credential or silently burns a student's allocation has failed at
something more important than pedagogy.

\section{The Diagnostic Learning Layer}
\label{sec:learning}

The compilation engine is the technical core, but a tool that only experts can
reach is of limited use. The learning layer places a curriculum in front of it:
ten modules running from the bit--qubit distinction to compiling a circuit and
executing it on hardware, over a dependency graph of $33$ topics to which $37$
quiz items and $13$ circuit-construction challenges are attached.

\subsection{Grading by simulation}

A challenge states a goal --- produce a given state, or a given measurement
distribution --- optionally with constraints such as a gate budget. Submissions
are graded by \emph{running} the circuit and comparing against the goal up to
global phase. No reference solution is compared against. The consequence is
that any circuit achieving the goal passes, including ones the authors did not
anticipate, which is both correct and necessary: the alternative, matching a
stored answer, teaches that there is one right circuit, which is false.

\subsection{Mastery from weighted evidence}

Each piece of evidence contributes a weight $w_{i}$ and a value
$v_{i} \in [0,1]$: a passed challenge $(2.0, 1.0)$, a failed challenge
$(1.0, 0.0)$, a quiz item $(1.0, 0$ or $1)$, a completed lesson $(0.5, 0.6)$.
For a topic with evidence set $I$,
\begin{equation}
\mu = \frac{\sum_{i \in I} w_{i} v_{i}}{\sum_{i \in I} w_{i}},
\qquad
c = \min\Bigl(1, \tfrac{1}{4}\textstyle\sum_{i \in I} w_{i}\Bigr),
\label{eq:mastery}
\end{equation}
where, in \eqref{eq:mastery}, $\mu$ is mastery and $c$ a confidence that reflects how much evidence
exists. A topic is \emph{mastered} when $\mu \ge 0.8$ and $c \ge 0.6$;
\emph{needs revision} when $\mu < 0.6$ on non-trivial evidence; \emph{locked}
when a prerequisite is below $0.6$; and \emph{ready} otherwise
(Table~\ref{tab:mastery}).

\begin{table}[t]
\caption{The whole mastery model. Four evidence weights and four thresholds,
all fixed and published; there are no learned parameters.}
\label{tab:mastery}
\centering
\setlength{\tabcolsep}{4pt}
\footnotesize
\begin{tabular}{@{}lrr@{}}
\toprule
Evidence & Weight $w_{i}$ & Value $v_{i}$ \\
\midrule
Challenge passed      & 2.0 & 1.0 \\
Challenge failed      & 1.0 & 0.0 \\
Quiz item             & 1.0 & 0 or 1 \\
Lesson completed      & 0.5 & 0.6 \\
\midrule
\multicolumn{3}{@{}l@{}}{\emph{Topic state}} \\
Mastered        & \multicolumn{2}{r@{}}{$\mu \ge 0.8$ and $c \ge 0.6$} \\
Needs revision  & \multicolumn{2}{r@{}}{$\mu < 0.6$ on non-trivial evidence} \\
Locked          & \multicolumn{2}{r@{}}{a prerequisite below $0.6$} \\
Ready           & \multicolumn{2}{r@{}}{otherwise} \\
\bottomrule
\end{tabular}
\end{table}

We score a challenge twice as heavily as a quiz because building a working
circuit shows more understanding than picking the right answer. Finished
lessons count for $0.6$ rather than a full point, because reading exposes you
to information but does not mean you can use it. These are design choices,
spelled out plainly so that people can disagree with them or adjust them. We
stick with a transparent rule rather than a learned
one~\cite{corbett1995knowledge, piech2015deep}: every recommendation carries a
clear justification, and a black-box network with hidden weights could never
provide that.

\subsection{Diagnosis by rewriting}

We are interested in \emph{why} an attempt failed. Rather than trying to infer
intent from a learner's text, the system rewrites the submitted circuit
mechanically in three specific ways and reruns it:

\begin{itemize}
\item exchange control and target on every CNOT;
\item apply each transposition of the circuit's wires;
\item compare against variants that add a missing Hadamard, or correct a
  phase sign.
\end{itemize}

If any transformation gets the circuit to meet the goal, that is the
diagnosis, and the system reports it complete with the evidence --- ``your
CNOTs are the right way round in principle but reversed here'', together with
the fact that reversing them produces the target state. If nothing fixes it,
the attempt is simply marked incorrect and the relevant lesson offered. The
system never invents a diagnosis. The rewrites test for seven common
misconceptions, always in a set order so that more specific issues take
precedence.

One safeguard is essential, and we added it only after spotting its absence: a
circuit that already works is never ``diagnosed'', whatever its structure. The
Bell state $(\ket{00} + \ket{11})/\sqrt{2}$ is symmetric under swapping its
two qubits, so a wire-permutation detector would happily flag a wrong-qubit
mistake in a perfectly correct submission. Correctness therefore always comes
first, and only a failing circuit meets the rewrite battery --- unless there is
a constraint violation, such as exceeding the gate budget, which can make
sense even when the outcome is right.

\subsection{Separation of scoring concerns}

Practice and the platform score different things, and an earlier version of the
system reported both as one. It therefore described the same learner as being
at one level on the practice page, where only challenge and quiz results
counted, and at a different level on the dashboard, where completed lessons
counted too. Here is the rule now: the platform controls experience level and
mastery, while practice sections only track their own results. Practice links
out to the main dashboard instead of showing a score of its own --- because as
soon as two places compute a score differently, someone ends up seeing two
conflicting numbers, and that is a recipe for confusion. We record it not
because the fix is interesting but because the failure is systematic.

\section{Evaluation}
\label{sec:evaluation}

\subsection{Setup}
\label{sec:setup}

All experiments ran on an AMD Ryzen 7 7445HS, eight cores and sixteen
threads, with 15.2\,GiB of RAM, Windows 11, CPython 3.12.9, Qiskit 2.5.0, Aer
0.17.2 and NumPy 2.2.6. We fixed the random seed at $42$. Every experiment is
driven by a script that writes its results as JSON, and every table in this
section is generated from those files automatically --- no measured number
here was copied across by hand.

\begin{table*}[t]
\caption{Inserted SWAP gates by router, device and benchmark circuit; lower is better.}
\label{tab:swaps}
\centering
\setlength{\tabcolsep}{3.5pt}
\footnotesize
\begin{tabular}{@{}llrrrrrr@{}}
\toprule
Device & Router & GHZ\mbox{-}7 & QFT\mbox{-}6 & CX\mbox{-}chain\mbox{-}7 & QAOA\mbox{-}ring\mbox{-}7 & Toffoli\mbox{-}5 & Random\mbox{-}7\mbox{-}d8 \\
\midrule
Linear-7 & \textbf{Qyantram} & 5 & 20 & 0 & 5 & 11 & 39 \\
 & Qiskit SABRE & 6 & 13 & 0 & 5 & 7 & 31 \\
 & BasicSwap & 5 & 40 & 0 & 5 & 11 & 37 \\
 & Lookahead & 5 & 20 & 0 & 5 & 6 & \emph{t/o} \\
\midrule
Ring-7 & \textbf{Qyantram} & 5 & 17 & 0 & 0 & 11 & 20 \\
 & Qiskit SABRE & 6 & 13 & 0 & 0 & 7 & 17 \\
 & BasicSwap & 5 & 23 & 0 & 0 & 11 & 26 \\
 & Lookahead & 5 & 16 & 0 & 0 & 6 & 26 \\
\midrule
Grid-9 & \textbf{Qyantram} & 4 & 9 & 12 & 5 & 12 & 23 \\
 & Qiskit SABRE & 3 & 7 & 5 & 7 & 8 & 15 \\
 & BasicSwap & 5 & 13 & 6 & 6 & 9 & 32 \\
 & Lookahead & 3 & 7 & 5 & 4 & 7 & 17 \\
\midrule
Heavy-hex-19 & \textbf{Qyantram} & 11 & 21 & 21 & 20 & 23 & 63 \\
 & Qiskit SABRE & 13 & 22 & 20 & 15 & 18 & 36 \\
 & BasicSwap & 13 & 44 & 43 & 20 & 22 & 79 \\
 & Lookahead & 17 & 21 & 12 & 14 & 15 & \emph{t/o} \\
\midrule
Eagle-127 & \textbf{Qyantram} & 5 & 20 & 0 & 5 & 11 & 39 \\
 & Qiskit SABRE & 6 & 13 & 0 & 5 & 7 & 31 \\
 & BasicSwap & 5 & 40 & 0 & 5 & 11 & 37 \\
 & Lookahead & 5 & 27 & 0 & 6 & \emph{t/o} & \emph{t/o} \\
\midrule
Heron-156 & \textbf{Qyantram} & 5 & 20 & 0 & 5 & 11 & 39 \\
 & Qiskit SABRE & 6 & 13 & 0 & 5 & 7 & 31 \\
 & BasicSwap & 5 & 40 & 0 & 5 & 11 & 37 \\
 & Lookahead & 5 & 27 & 0 & 6 & 7 & \emph{t/o} \\
\bottomrule
\end{tabular}
\\[3pt]
\begin{minipage}{0.92\textwidth}\footnotesize
\emph{t/o} marks a router that did not finish inside the 60\,s per-measurement budget. Our router starts from the identity layout and performs no optimisation, which is the cost of emitting a decision record; Qiskit's routers run with a trivial layout and \texttt{optimization\_level=0}, so the comparison isolates routing.
\end{minipage}
\end{table*}

\subsection{Routing quality, and the cost of transparency}
\label{sec:quality}

We route six benchmark circuits --- a 7-qubit GHZ preparation, a 6-qubit QFT, a
three-layer CNOT chain, a 7-qubit QAOA ring layer, a chain of three Toffoli
gates, and a depth-8 random two-qubit circuit, all reduced to one- and
two-qubit gates --- onto six device graphs: linear (7 qubits, 6 edges), ring
(7, 7), grid (9, 12), heavy-hex $d=3$ (19, 20), the Eagle layout (127, 144) and
the Heron~r2 layout (156, 176). That is 36 device/circuit pairs. Each pair is
routed by our router and by Qiskit's SABRE, BasicSwap and LookaheadSwap, giving
144 measurements. Each measurement runs in a child process under a 60\,s
budget. For a routing-only comparison, the Qiskit routers are invoked with a
trivial (identity) layout and \texttt{optimization\_level=0}, matching our
router's identity start and absence of optimisation.

Table~\ref{tab:swaps} gives the SWAP counts and Table~\ref{tab:aggregate} the
aggregate. The result is that our router is worse than Qiskit's SABRE, by a
mean factor of $1.27$ in SWAP count and $1.10$ in depth. Across the 36 comparison pairs our routing needed fewer SWAPs on 7, matched
Qiskit's count on 8, and used more on 21, with per-pair ratios from $0.71$ to
$2.40$ and a median of $1.28$. BasicSwap is worse again
at $1.53\times$; LookaheadSwap attains $1.08\times$ but failed to finish within
the budget on 5 of 36 pairs, including on the 7-qubit linear chain, which is
the smallest device in the study.

This extra cost is not a problem with SABRE --- it is the result of our
design choice. We always start from the identity layout with no optimisation,
so every assignment of logical to physical qubits can be tracked step by step:
a reader can follow $q_{i} \mapsto p_{i}$ through every round. Qiskit's SABRE,
by contrast, runs an extra layout-selection phase that routes the circuit
forwards and backwards several times to pick a good initial mapping, and that
forward-and-back tweaking is mostly responsible for the SWAP gap. In sum: our
identity start costs about $27\%$ more SWAPs, but gives a routing record that
is easy to follow and audit.

\begin{table}[t]
\caption{Aggregate over every device/circuit pair that completed. Ratios are per-pair against Qiskit SABRE on the same pair, then averaged.}
\label{tab:aggregate}
\centering
\setlength{\tabcolsep}{3.5pt}
\footnotesize
\begin{tabular}{@{}lrrrrr@{}}
\toprule
Router & $n$ & t/o & $\Sigma$ SWAP & SWAP ratio & depth ratio \\
\midrule
\textbf{Qyantram} & 36 & 0 & 517 & 1.27$\times$ & 1.10$\times$ \\
Qiskit SABRE & 36 & 0 & 398 & 1.00$\times$ & 1.00$\times$ \\
BasicSwap & 36 & 0 & 651 & 1.53$\times$ & 1.16$\times$ \\
Lookahead & 31 & 5 & 294 & 1.08$\times$ & 1.03$\times$ \\
\bottomrule
\end{tabular}
\end{table}

\begin{table}[t]
\caption{Runtime, and the size of the decision record emitted.}
\label{tab:cost}
\centering
\setlength{\tabcolsep}{3.5pt}
\footnotesize
\begin{tabular}{@{}lrrrr@{}}
\toprule
Router & mean (ms) & max (ms) & rounds & scored cands. \\
\midrule
\textbf{Qyantram} & 2.6 & 9.4 & 64 & 521 \\
Qiskit SABRE & 1432.1 & 2275.0 & 0 & 0 \\
BasicSwap & 1438.0 & 2335.3 & 0 & 0 \\
Lookahead & 1469.3 & 2513.6 & 0 & 0 \\
\bottomrule
\end{tabular}
\\[3pt]
\begin{minipage}{\columnwidth}\footnotesize
Our figure is the routing call alone, \emph{including} building the full decision record. The Qiskit figures are end-to-end \texttt{transpile()} calls and so also contain pass-manager and target construction; they bound routing time from above and are \textbf{not} like-for-like, so no speed advantage is claimed here. The last two columns are the point: the record costs single-digit milliseconds.
\end{minipage}
\end{table}

\subsection{Cost of the decision record}
\label{sec:cost}

Table~\ref{tab:cost} reports what the record costs to produce. Over all 36
pairs the routing call, \emph{including} construction of the complete record,
took under 10\,ms in the worst case and a few milliseconds on average; the
largest record covered 64
rounds and 521 scored candidates. On the two device-scale graphs (127 and 156
qubits) the worst case was under 7\,ms and 256 scored candidates. Recording every
candidate SWAP with its decomposed score is therefore a few milliseconds, not a
change of complexity class --- which is the result that matters, since it means
transparency need not be an optional mode.

Determinism was checked by re-routing every pair three further times and
comparing the SWAP count, the routed depth and the final mapping. All 36 pairs
produced byte-identical results on all repeats, as the tie-break on the qubit
pair requires. This matters for a teaching tool: a learner who re-runs an
example must see the same decisions.

\begin{table}[t]
\caption{Equivalence-check cost for one fixed 5-qubit circuit as the device grows. The restricted check is independent of device size because it never builds the idle wires. Best of 9 runs.}
\label{tab:verify}
\centering
\setlength{\tabcolsep}{3.5pt}
\footnotesize
\begin{tabular}{@{}rrrrr@{}}
\toprule
$N$ & $2^{N}$ amplitudes & unrestricted & restricted & speed-up \\
\midrule
5 & 32 & 0.84\,ms & 0.828\,ms & 1.0$\times$ \\
8 & 256 & 0.94\,ms & 0.845\,ms & 1.1$\times$ \\
10 & 1\,024 & 1.19\,ms & 0.860\,ms & 1.4$\times$ \\
12 & 4\,096 & 1.86\,ms & 0.817\,ms & 2.3$\times$ \\
14 & 16\,384 & 6.40\,ms & 0.828\,ms & 7.7$\times$ \\
16 & 65\,536 & 54.86\,ms & 0.848\,ms & 65$\times$ \\
18 & 262\,144 & 135.09\,ms & 0.672\,ms & 201$\times$ \\
20 & 1\,048\,576 & 528.02\,ms & 0.563\,ms & 938$\times$ \\
22 & 4\,194\,304 & 2302.37\,ms & 0.937\,ms & 2\,456$\times$ \\
24 & 16\,777\,216 & 10036.14\,ms & 0.921\,ms & 10\,895$\times$ \\
\midrule
127 & $1.7\times 10^{38}$ & $2.5\times 10^{30}$\,GiB & 0.828\,ms & --- \\
156 & $9.1\times 10^{46}$ & $1.4\times 10^{39}$\,GiB & 0.828\,ms & --- \\
\bottomrule
\end{tabular}
\\[2pt]\footnotesize The last two rows are the IBM Eagle and Heron r2 layouts the engine models; the unrestricted check is not merely slow there but impossible.
\end{table}

\subsection{Verification cost}

Table~\ref{tab:verify} measures the reduction of
Section~\ref{sec:verification}. A fixed, fully entangling 5-qubit circuit is
embedded in a device register of growing size $N$, and we time the state-vector
construction that the equivalence check needs, unrestricted and restricted to
the occupied wires, taking the best of nine runs.

The unrestricted cost grows as expected --- under a millisecond at $N=5$
rising to about ten seconds at $N=24$, roughly a factor of four per two qubits
added --- while the restricted cost stays under a millisecond across the whole
range, because the idle wires are never built. The speed-up therefore grows
from unity to around four orders of magnitude over the range measured.
Table~\ref{tab:verify} gives the figures from a single run.

A caution about those figures, since they are the only numbers in this paper
that do not reproduce exactly. The restricted check is sub-millisecond, which
is close enough to scheduler granularity that ordinary background load on the
machine perturbs it. Repeating the whole experiment on the same host gave
speed-ups at $N=24$ between roughly $6\,000\times$ and $14\,000\times$,
depending on what else was running, even with the best-of-nine minimum. The
order of magnitude is stable and the shape of the two curves --- one
exponential, one flat --- is not in question, but the precise ratio should be
read as approximate. Every other quantity reported in this paper is computed
by algebra or exhaustive enumeration and is reproducible exactly.

The last two rows of Table~\ref{tab:verify} are the point. On the Eagle and
Heron layouts the engine actually models, the unrestricted comparison would
need $1.7 \times 10^{38}$ and $9.1 \times 10^{46}$ amplitudes, requiring on the
order of $10^{30}$ and $10^{39}$ GiB. The restricted comparison for the same
5-qubit circuit needs $32$ amplitudes and under a millisecond. The difference is not
one of degree: without the reduction, verification of a compilation targeting a
real device is not slow but impossible, and so in practice would simply not be
done.

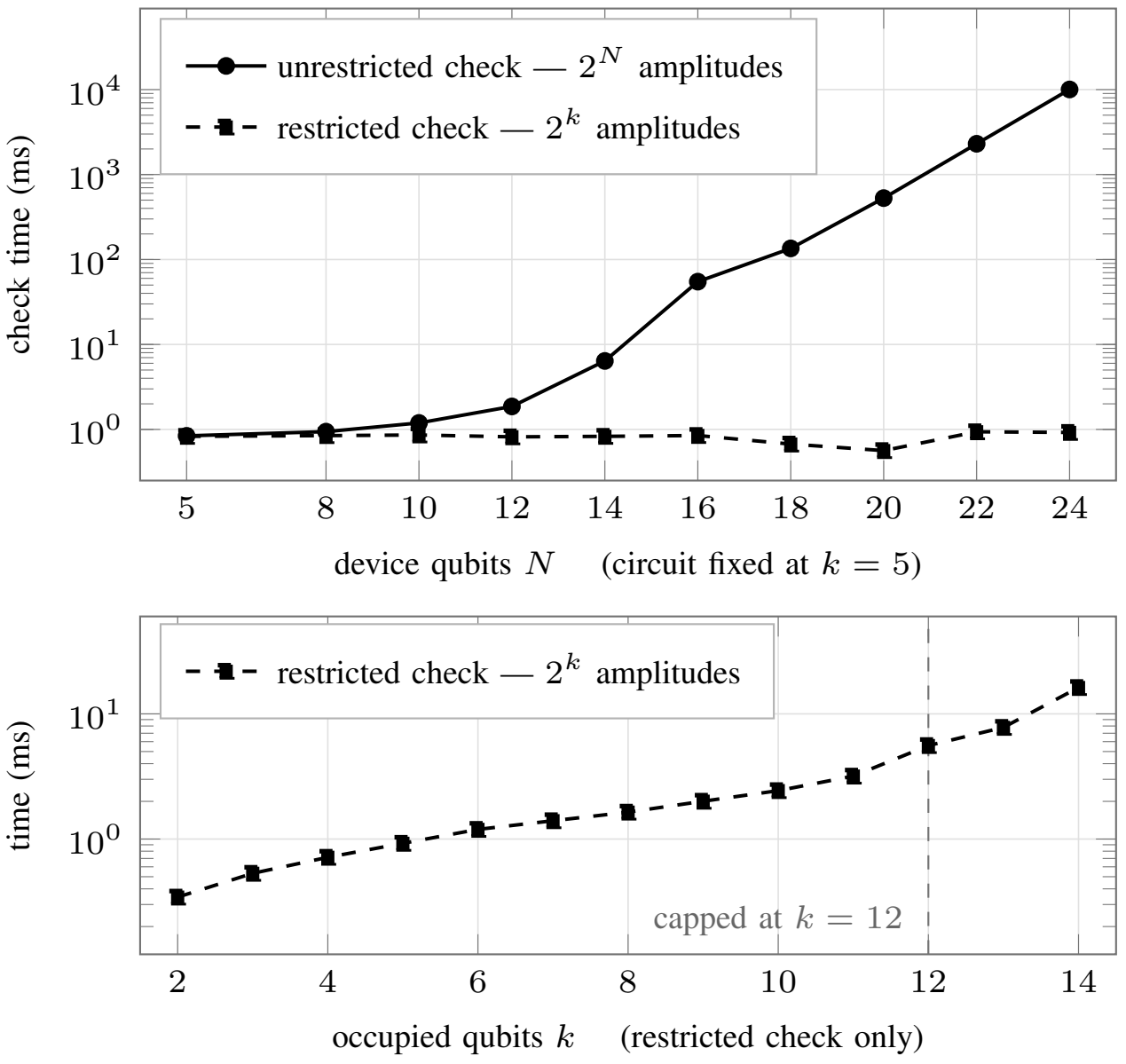
\begin{figure}[t]
\centering
\pgfplotsset{
  verifyaxis/.style={
    width=\columnwidth,
    tick label style={font=\scriptsize},
    label style={font=\scriptsize},
    grid=major,
    major grid style={black!12, line width=0.3pt},
    axis line style={black!55},
    tick style={black!55},
    every axis plot/.append style={line width=0.75pt},
  },
}
\begin{tikzpicture}
\begin{semilogyaxis}[
  verifyaxis,
  height=5.1cm,
  xlabel={device qubits $N$ \quad(circuit fixed at $k=5$)},
  ylabel={check time (ms)},
  xmin=4, xmax=25,
  ymin=0.25, ymax=90000,
  xtick={5,8,10,12,14,16,18,20,22,24},
  ytick={1,10,100,1000,10000},
  legend style={font=\scriptsize, at={(0.02,0.98)}, anchor=north west,
                draw=black!30, fill=white, row sep=1pt, inner xsep=5pt,
                inner ysep=4pt},
  legend cell align=left,
]
\addplot[mark=*, mark size=1.5pt, black]
  table[x=N, y=full] {tables/verify.dat};
\addlegendentry{unrestricted check --- $2^{N}$ amplitudes}
\addplot[mark=square*, mark size=1.4pt, black, dashed]
  table[x=N, y=reduced] {tables/verify.dat};
\addlegendentry{restricted check --- $2^{k}$ amplitudes}
\end{semilogyaxis}
\end{tikzpicture}

\vspace{4pt}

\begin{tikzpicture}
\begin{semilogyaxis}[
  verifyaxis,
  height=4.1cm,
  xlabel={occupied qubits $k$ \quad(restricted check only)},
  ylabel={time (ms)},
  xmin=1.5, xmax=14.5,
  ymin=0.12, ymax=60,
  xtick={2,4,6,8,10,12,14},
  ytick={0.1,1,10},
  legend style={font=\scriptsize, at={(0.02,0.98)}, anchor=north west,
                draw=black!30, fill=white, inner xsep=5pt, inner ysep=4pt},
  legend cell align=left,
]
\addplot[mark=square*, mark size=1.4pt, black, dashed]
  table[x=k, y=seconds] {tables/occupancy.dat};
\addlegendentry{restricted check --- $2^{k}$ amplitudes}
\draw[black!55, dashed] (axis cs:12,0.12) -- (axis cs:12,60);
\node[font=\scriptsize, anchor=south east, black!60, inner sep=2pt]
  at (axis cs:11.8,0.14) {capped at $k=12$};
\end{semilogyaxis}
\end{tikzpicture}

\caption{Measured equivalence-check cost (log scale). \textbf{Top:} with the
circuit fixed at five qubits and the device growing, the unrestricted check
grows exponentially while the restricted one stays flat --- it never builds
the idle wires. \textbf{Bottom:} the restricted check's own growth is in
circuit \emph{width}, not device size, which is why the engine caps it at
$k=12$.}
\label{fig:verifyplot}
\end{figure}

Fig.~\ref{fig:verifyplot} plots both curves, together with the restricted
check's own growth in circuit width $k$, which is what the $k \le 12$ ceiling
reflects: a few milliseconds at $k=12$, several times that by $k=14$.

%

\subsection{Error-identification validation}
\label{sec:qecresults}

Table~\ref{tab:qecvalid} compares the module's enumerated logical error rate
for the three-qubit bit-flip code under independent $X$ noise against the closed
form $3p^{2} - 2p^{3}$, at eleven noise strengths from $p = 10^{-3}$ to
$p = 0.5$. The largest absolute deviation across both repetition codes and all
eleven values is $2.8 \times 10^{-17}$, i.e.\ agreement at the level of
double-precision rounding. This is the expected outcome and that is the point:
because the module enumerates every error pattern rather than sampling,
disagreement with algebra would indicate a defect in the implementation, so the
comparison functions as a test oracle rather than a statistical validation.

Table~\ref{tab:qeccost} reports the cost. The nine-qubit Shor code has
$4^{9} = 262\,144$ error patterns and $2^{8} = 256$ distinct syndromes; the
NumPy-powered full enumeration finishes in a fraction of a second, which
makes exact logical error rates practical within normal web
latency. Pseudo-thresholds come from bisection on the enumerated curve.

\begin{figure}[t]
\centering
\begin{tikzpicture}
\begin{loglogaxis}[
  width=\columnwidth, height=6.0cm,
  xlabel={physical error rate $p$ (depolarising)},
  ylabel={logical error rate $p_{L}$},
  xmin=8e-4, xmax=0.6,
  ymin=1e-6, ymax=4,
  tick label style={font=\scriptsize},
  label style={font=\scriptsize},
  grid=major,
  major grid style={black!12, line width=0.3pt},
  axis line style={black!55},
  tick style={black!55},
  legend style={font=\scriptsize, at={(0.98,0.03)}, anchor=south east,
                draw=black!30, fill=white, row sep=1.5pt,
                inner xsep=5pt, inner ysep=4pt},
  legend cell align=left,
  every axis plot/.append style={line width=0.75pt},
]

\addplot[black!50, dotted, line width=1pt]
  coordinates {(8e-4,8e-4) (0.6,0.6)};
\addlegendentry{break-even, $p_{L}=p$}

\addplot[mark=*, mark size=1.4pt, black]
  table[x=p, y=logical] {tables/qec-bit_flip.dat};
\addlegendentry{repetition codes, 3 qubits}

\addplot[mark=triangle*, mark size=1.7pt, black, dashdotted]
  table[x=p, y=logical] {tables/qec-shor.dat};
\addlegendentry{Shor code, 9 qubits}

\end{loglogaxis}
\end{tikzpicture}
\caption{Logical error rate against physical error rate under depolarising
noise, from exhaustive enumeration of all $4^{n}$ error patterns rather than
sampling. Encoding helps only where a curve lies \emph{below} the dotted
break-even line. Shor's nine-qubit code does, up to $p^{*}=0.085$. The
bit-flip and phase-flip codes do not, at any $p$ measured: each sits at
$p_{L}\approx 2p$, adding more exposure than it removes, because each
corrects one Pauli type and is blind to the other two. That symmetry also
makes their two curves identical to within $6\times10^{-17}$, so they are
drawn as one. Plotting the codes that fail is deliberate --- showing only the
one that works would teach that encoding is always an improvement.}
\label{fig:qecplot}
\end{figure}
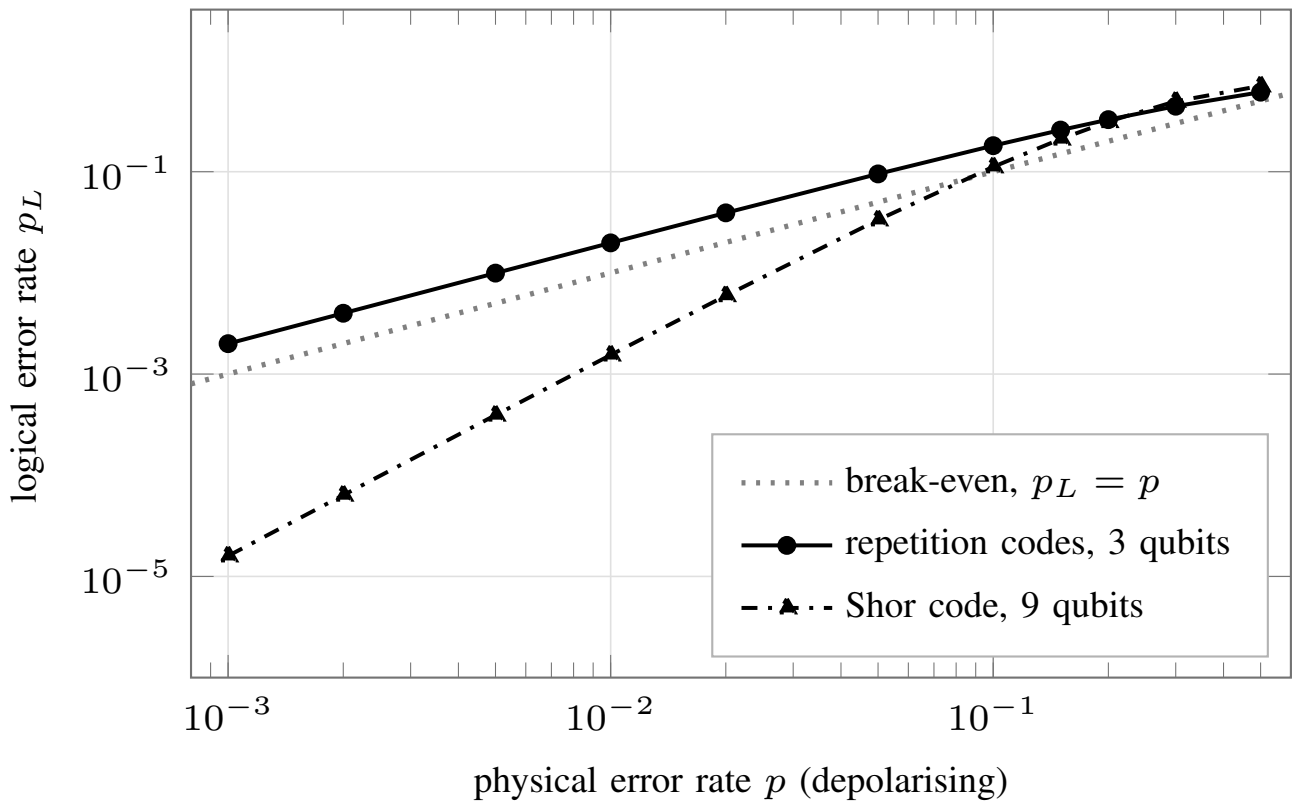

The thresholds teach an important lesson. With the bit-flip code under matched
noise you always do better, since $3p^{2} - 2p^{3} < p$ at any $p < 0.5$.
Under depolarising noise Shor's code helps up to $p^{*} = 0.085$, but plain
repetition codes help nowhere: each corrects a single Pauli type and is blind
to the other two, so the logical error rate sticks at $p_{L} \approx 2p$,
always worse than leaving the qubit unencoded. Fig.~\ref{fig:qecplot} shows
this against the break-even line $p_{L} = p$. The area above that line is
exactly what gets hidden if a module only reports successful corrections ---
and it is why the full nine-qubit Shor structure matters.

\begin{table}[t]
\caption{Exhaustive enumeration against closed form, three-qubit bit-flip code under independent $X$ noise.}
\label{tab:qecvalid}
\centering
\setlength{\tabcolsep}{3.5pt}
\footnotesize
\begin{tabular}{@{}rlll@{}}
\toprule
$p$ & enumerated & $3p^{2}-2p^{3}$ & $|\Delta|$ \\
\midrule
0.001 & 0.000002998000 & 0.000002998000 & $4.2\times 10^{-22}$ \\
0.002 & 0.000011984000 & 0.000011984000 & $0$ \\
0.005 & 0.000074750000 & 0.000074750000 & $1.4\times 10^{-20}$ \\
0.01 & 0.000298000000 & 0.000298000000 & $0$ \\
0.02 & 0.001184000000 & 0.001184000000 & $2.2\times 10^{-19}$ \\
0.05 & 0.007250000000 & 0.007250000000 & $0$ \\
0.1 & 0.028000000000 & 0.028000000000 & $0$ \\
0.15 & 0.060750000000 & 0.060750000000 & $6.9\times 10^{-18}$ \\
0.2 & 0.104000000000 & 0.104000000000 & $1.4\times 10^{-17}$ \\
0.3 & 0.216000000000 & 0.216000000000 & $2.8\times 10^{-17}$ \\
0.5 & 0.500000000000 & 0.500000000000 & $0$ \\
\bottomrule
\end{tabular}
\\[3pt]
\begin{minipage}{\columnwidth}\footnotesize
Every error pattern is enumerated rather than sampled, so agreement is exact to floating-point precision. The largest deviation across both repetition codes and all noise strengths tested is $2.8\times 10^{-17}$.
\end{minipage}
\end{table}

\begin{table}[t]
\caption{Enumeration cost, and the pseudo-threshold $p^{*}$ below which encoding reduces the error rate.}
\label{tab:qeccost}
\centering
\setlength{\tabcolsep}{3.5pt}
\footnotesize
\begin{tabular}{@{}lrrrll@{}}
\toprule
Code & patterns & syndromes & time & $p^{*}$ ($X$) & $p^{*}$ (depol.) \\
\midrule
Bit-flip & 64 & 4 & 0.1\,ms & 0.500 & \emph{none} \\
Phase-flip & 64 & 4 & 0.1\,ms & \emph{none} & \emph{none} \\
Shor-9 & 262\,144 & 256 & 311.2\,ms & 0.135 & 0.085 \\
\bottomrule
\end{tabular}
\\[3pt]
\begin{minipage}{\columnwidth}\footnotesize
\emph{none} means the encoded rate exceeds the bare rate at every $p$: a repetition code is blind to one of the two error types, so under depolarising noise it adds more exposure than it removes. That is a property of the codes, not of this implementation, and it is why Shor's code exists.
\end{minipage}
\end{table}

%
%

\subsection{The machine-learning engine against a general simulator}
\label{sec:qmlresults}

Section~\ref{sec:qml} argues that the training loop needs its own state-vector
engine. Table~\ref{tab:qml} tests both halves of that argument.

Correctness first: over every circuit width from one to ten qubits and depths
of one, three and six, the engine's output is compared with
\texttt{qiskit.quantum\_info} up to global phase. The worst infidelity is
$2.0 \times 10^{-15}$ --- agreement at double-precision rounding, as it must
be, since the two are computing the same product of the same unitaries in a
different order of operations.

Speed second, and here the distinction that matters is between one circuit and
a batch. On a single circuit the engine is between roughly $1.7\times$ and $2.6\times$
faster, which would not on its own justify maintaining a second simulator. On
the batched path --- one circuit applied to every sample, which is what one
training iteration is --- it is roughly $25\times$ to $50\times$ faster, because the
per-call work is paid once for the batch instead of once per sample. That is
the difference between a parameter-shift gradient being feasible inside a web
request and not, and it is the reason the duplication is there.

\begin{table}[t]
\caption{The QML Studio state-vector engine against \texttt{qiskit.quantum\_info}.}
\label{tab:qml}
\centering
\setlength{\tabcolsep}{3.5pt}
\footnotesize
\begin{tabular}{@{}llrrr@{}}
\toprule
Mode & Size & Qyantram & Qiskit & Speed-up \\
\midrule
single & 4 qubits & 300\,$\mu$s & 779\,$\mu$s & 2.6$\times$ \\
single & 6 qubits & 499\,$\mu$s & 1\,189\,$\mu$s & 2.4$\times$ \\
single & 8 qubits & 795\,$\mu$s & 1\,796\,$\mu$s & 2.3$\times$ \\
single & 10 qubits & 1\,374\,$\mu$s & 2\,829\,$\mu$s & 2.1$\times$ \\
single & 12 qubits & 2\,965\,$\mu$s & 4\,938\,$\mu$s & 1.7$\times$ \\
\midrule
batch & 16 states & 0.62\,ms & 16.07\,ms & \textbf{26$\times$} \\
batch & 64 states & 1.24\,ms & 64.95\,ms & \textbf{52$\times$} \\
batch & 256 states & 5.52\,ms & 280.73\,ms & \textbf{51$\times$} \\
\bottomrule
\end{tabular}
\\[3pt]
\begin{minipage}{\columnwidth}\footnotesize
\emph{Single} is one circuit evaluated once; \emph{batch} is one circuit applied to many input states at once, which is what a training iteration does. The two engines agree to a worst infidelity of $2.0\times 10^{-15}$ over every width and depth tested.
\end{minipage}
\end{table}

\subsection{Software validation}

\qy{} includes 22 test modules and 6\,391 lines of tests: 599 cases, with just
one skip, running in 75.6\,s. These are not for show --- they enforce the
claims above. Transpiler tests check the router's output against the reference
implementation it was ported from and verify that the documented heuristic
constants have not drifted. The error-identification logic is validated
against independent Pauli algebra, for both single- and multi-qubit errors,
and each correction is tested by state fidelity. Machine-learning tests rerun
the cross-check of Section~\ref{sec:qmlresults} after every code change, so
drift between the two engines is caught right away rather than quietly
changing a model's predictions. And one specific check walks the compilation
engine's transitive imports and fails if any hardware-submission logic sneaks
in --- which is how the isolation property of
Section~\ref{sec:architecture} is kept true rather than merely intended.

\section{Discussion and Limitations}
\label{sec:discussion}

\subsection{What the measurements support, and what they do not}

Four claims in this paper are supported by the measurements reported above.
The decision record costs a few milliseconds and does not change the
complexity class of routing (Table~\ref{tab:cost}). The occupied-sub-space
reduction makes per-compilation equivalence verification independent of device
size, turning an impossible check into a sub-millisecond one on the device
layouts the engine models (Table~\ref{tab:verify}). The
error-identification module agrees with closed-form algebra to within
$2.8 \times 10^{-17}$ (Table~\ref{tab:qecvalid}). And the machine-learning
engine agrees with \texttt{qiskit.quantum\_info} to
$2.0 \times 10^{-15}$ while running the batched training path roughly $25$--$50\times$
faster (Table~\ref{tab:qml}).

Several claims are \emph{not} supported, and we would rather say so than let
them be inferred.

\subsubsection*{We do not improve routing quality}
Our method uses about $1.27\times$ as many SWAPs as Qiskit's SABRE
(Table~\ref{tab:aggregate}). That is intentional: the contribution is
transparency, not winning a SWAP-count contest. Anyone focused purely on
minimising execution cost should use a production compiler. Combining
transparency and quality is not impossible --- you could run a layout search
and still keep a readable decision record --- and bringing those together is
the next obvious step.

\subsubsection*{Verification checks one input, not unitary equality}
Our verifier compares only what the circuit does to the
$\ket{0}^{\otimes N}$ input, not whether the entire unitaries are identical.
If an error leaves that input unchanged, it gets through.
Decision-diagram~\cite{burgholzer2021advanced} and
path-sum~\cite{amy2018towards} approaches check full unitary equivalence and
are therefore stronger. Still, our method is orthogonal and could work
alongside them: you could limit a decision-diagram check to the occupied
sub-space, just as we do, to shrink the problem.

\subsubsection*{The circuit-width ceiling is real}
The width ceiling removes scaling with device size $N$, but not with circuit
width $k$. At $k = 12$ the check still costs a few milliseconds; by $k = 14$
it is several times that, and after that things blow up exponentially. We still compile wide
circuits and return them, but verification is skipped and that is spelled out
in the response. Extending coverage means abandoning state vectors for a
representation that does not enumerate amplitudes.

\subsubsection*{No learning outcomes are measured}
This is the most important limitation. Section~\ref{sec:learning} describes an
architecture and validates its mechanism: the mastery rule is deterministic and
inspectable, and the misconception diagnoses are confirmed by re-simulation
rather than guessed. Whether any of it improves what learners understand is an
empirical question this paper does not answer. We have not run a controlled
study, we have no cohort data, and we therefore make no claim about
educational effect. Designing that study --- pre/post assessment against a
control using conventional materials, with the diagnostic feedback as the
manipulated variable --- is the work we consider most urgent, and until it is
done the learning layer should be read as a hypothesis with a careful
implementation, not as a validated intervention.

\subsubsection*{Simulator scope}
The compilation engine is deliberately simulator-side and has no path to
hardware submission (Section~\ref{sec:architecture}). Consequently nothing in
this paper reports what happens when a compiled circuit meets a real device's
noise. The compilation is verified; the execution is not characterised.

\subsection{Design positions worth defending}

Two decisions recur through the system and are worth stating as positions
rather than details.

\emph{Report the losing candidates.} A system that shows only its conclusion
cannot be checked. Our router reports every candidate SWAP it rejected and the
score it rejected it at; the learning layer reports the reason behind every
recommendation; the error module reports mis-corrections and undetected errors
alongside successes. In each case the extra information is what makes the
output falsifiable by its reader.

\emph{Decline rather than degrade.} Every limit in the engine is explicit and
every refusal carries its reason: the equivalence check says it was skipped
because the circuit measures, or because it occupies thirteen qubits; the
topology builder rejects a disconnected graph by name rather than failing to
converge. In a teaching tool a clear refusal is more instructive than a silent
approximation, and in any tool it is easier to debug.

\subsection{Threats to validity}

The routing comparison uses six circuit families on six graphs with a single
seed; a broader benchmark suite, several seeds, and the randomised model
circuits of Cross~\textit{et al.}~\cite{cross2019validating} would give a
better estimate of the $1.27\times$ figure, which should be read as indicative
rather than precise. All timings come from one host and one software stack
(Section~\ref{sec:setup}); the verification speed-up is a ratio of two
measurements on the same host and so is robust to that, but the absolute
millisecond figures are not portable. Qiskit's routers were invoked through
\texttt{transpile()}, whose cost includes pass-manager construction, so
Table~\ref{tab:cost} does not license any claim that our router is faster, and
we make none.

\section{Conclusion}
\label{sec:conclusion}

Compilation is the point in the quantum stack where a circuit stops being the
object its author wrote, and it is conventionally the point at which the stack
stops explaining itself. We have described \qy{}, a platform built on the
position that this is a design choice rather than a necessity, and that
reversing it costs little.

Three mechanisms carry the argument. Separating decomposition, routing and
basis translation into stages that may each change exactly one property of the
circuit lets each added gate be traced to one cause, which turns ``my
circuit grew by a factor of thirty'' into two numbers with two different
remedies. Emitting the router's full decision record --- the blocked front
layer, the look-ahead set, and every candidate SWAP with its decomposed score,
including the ones rejected --- makes layout synthesis inspectable, and we
measure the cost at under 10\,ms and 521 scored candidates across 36
device/circuit pairs up to 156 qubits. Restricting equivalence verification to
the qubits a compiled circuit actually occupies makes the check exact,
independent of device size, and fast enough to run unconditionally: flat at
under a millisecond where the unrestricted comparison reaches seconds at 24 qubits
and $10^{30}$ GiB at 127.

The transparency is not free, and we have quantified rather than hidden the
price: with an identity initial layout and no optimisation, our router inserts
$1.27\times$ the SWAPs of Qiskit's SABRE. We regard that as a favourable
exchange for a teaching and debugging tool, and an unfavourable one for
production compilation --- which is why the two should be combined rather than
chosen between.

\subsection*{Future work}

Four directions follow directly. First, a layout-selection phase that still
emits a decision record, which should close most of the SWAP gap without
sacrificing observability. Second, applying the occupied-sub-space restriction
in front of a decision-diagram equivalence
checker~\cite{burgholzer2021advanced}, which would raise the width ceiling
while keeping independence from device size. Third, extending the
error-identification module to a surface-code patch, where the exhaustive
enumeration that validates the present codes becomes infeasible and must be
replaced by a sampled estimate with stated confidence intervals --- a change in
evidentiary standard that the module should be honest about. Fourth, and most
importantly, the controlled study of learning outcomes that
Section~\ref{sec:discussion} identifies as missing: without it, the
diagnostic layer remains a well-implemented hypothesis.

The platform, its 599 automated tests, and the four benchmark scripts that
produced every table in this paper are available as open source.

\section*{Reproducibility}
The four experiments in Section~\ref{sec:evaluation} are driven by the scripts
\texttt{bench\_routing.py}, \texttt{bench\_verify.py}, \texttt{bench\_qec.py}
and \texttt{bench\_qml.py}, each of which writes the JSON file from which the
corresponding results table is generated mechanically, so no measured value in
this paper was transcribed by hand. The tables that describe the system rather
than measure it --- implementation size, device graphs, model components,
evidence weights, and the illustrative routing record of
Table~\ref{tab:decision} --- are written out in the text, and are marked as
such where it matters. All randomness is seeded (seed $42$). Software versions
and host details are given in Section~\ref{sec:setup}.

\section*{Acknowledgment}
The authors thank the faculty of \dept{}, \inst{}, for access to computing
facilities and for guidance during this work.

\bibliographystyle{IEEEtran}
\bibliography{references}

\end{document}